\documentclass[twocolumn]{aastex631}

\usepackage{amsmath}	
\usepackage{amsfonts}
\usepackage{bm}
\usepackage{hyperref}
\usepackage[T1]{sansmath} 
\usepackage{graphicx}

\begin{document}

\title{Heavy element nucleosynthesis in rotating proto-magnetar winds}

\author{Tejas Prasanna}
\affiliation{Columbia Astrophysics Laboratory, Columbia University, New York, NY 10027, USA}

\author{Matthew S. B. Coleman}
\affiliation{Research Computing, Princeton University, Princeton, NJ 08544, USA}

\author[0000-0003-2377-9574]{Todd A. Thompson}
\affiliation{Center for Cosmology \& Astro-Particle Physics, The Ohio State University, Columbus, Ohio 43210, USA}
\affiliation{Department of Astronomy, The Ohio State University, Columbus, Ohio 43210, USA}
\affiliation{Department of Physics, The Ohio State University, Columbus, Ohio 43210, USA}

\author[0000-0002-4670-7509]{Brian D.~Metzger}
\affiliation{Department of Physics and Columbia Astrophysics Laboratory, Columbia University, New York, NY 10027, USA}
\affiliation{Center for Computational Astrophysics, Flatiron Institute, 162 5th Avenue, New York, NY 10010, USA}

\author[0009-0000-1335-4412]{Anirudh Patel}
\affiliation{Department of Physics and Columbia Astrophysics Laboratory, Columbia University, New York, NY 10027, USA}

\author{Bradley S. Meyer}
\affiliation{Department of Physics and Astronomy, Clemson University, Clemson, SC 29634, USA}

\begin{abstract}
The astrophysical origin of elements synthesized through the rapid neutron capture process ($r-$process) is a long standing mystery. The hot and dense environments of core-collapse supernovae have been suggested as potential $r-$process sites, particularly the neutrino-driven wind from the newly-born protoneutron star (PNS). Wind models that neglect the potential effects of strong magnetic fields and/or rapid rotation of the PNS typically fail to achieve the necessary conditions for production of the third $r-$process peak, but robustly produce a limited or weak $r-$process for neutron-rich winds. Axisymmetric magnetohydrodynamic simulations of rotating and non-rotating PNS winds with magnetar-strength fields reveal that high entropy material is quasi-periodically ejected from the equatorial closed zone of the PNS magnetosphere. Here, we post-process tracer particle trajectories from these simulations using a nuclear reaction network in order to explore the resulting nucleosynthesis across a range of PNS magnetic field strengths, rotation rates, and neutrino luminosities (cooling phase after core-bounce). We find that a robust $r-$process up to and beyond the third peak is generic to magnetar birth, even for magnetic fields as weak as $\sim 5\times 10^{14}$\,G. Depending on the distribution of magnetic field strengths and rotation at birth, we estimate that magnetized PNS winds could account for $\sim 5-100\%$ of the Galactic $r-$process inventory, extending up to the third peak. The robust $r-$process in our calculations is accompanied by overproduction of elements with mass number $\rm A\lesssim 120$ compared to the Solar abundances. We also find that $^{92}\rm Mo$ (a $p-$isotope) is produced in significant quantities in neutron-rich winds.        
\end{abstract}

\section{Introduction}
\label{intro}

The rapid neutron capture process ($r-$process) enriches our Universe with roughly half of the elements heavier than iron (e.g., \citealt{Burbidge1957}), yet its astrophysical origins remain debated.  Several potential sites of the $r-$process have been considered over the years, including neutron star mergers (e.g., \citealt{Lattimer1974, Eichler1989, Perego2021}), neutrino-driven winds that accompany the birth of protoneutron stars (PNSs) in the seconds following successful core-collapse supernova explosions \citep{QW1996, Otsuki2000, Thompson2001, Thompson2018}, accretion disk outflows in collapsars \citep{Siegel2019}, and magnetorotational core-collapse supernovae \citep{Nishimura2015}, among other sources. Although neutron star mergers are likely major if not dominant sources of $r-$process elements in our Galaxy (e.g., \citealt{Cote2018}), several studies indicate other sites are likely to be operating as well (e.g., \citealt{Qian2007, Sneden2008, Cote2019, Zevin2019, Simon2023, Ou2024}). Indeed, recent work found evidence in the form of gamma-ray radioactive decay emission that an $r-$process occurred in the ejecta from the giant flare of a Galactic magnetar  \citep{Cehula2024,Patel2025,Patel2025_2}.

Formed in the aftermath of supernova explosions, PNSs cool over several seconds through neutrino emission \citep{Burrows1986}. Heating of the PNS atmosphere by these neutrinos drives a thermal wind from their surfaces \citep{Duncan1986, Burrows1995, Janka1996}. Though long suggested as potential sites for producing heavy elements \citep{Woosley1992, Meyer1992, Woosley1994}, detailed PNS wind models found they do not achieve the conditions required for the production of the third $r-$process peak \citep{QW1996}, except for the most massive and compact PNSs at the highest neutrino luminosities achieved very early in the cooling phase \citep{Otsuki2000, Thompson2001, Wanajo2001}.

Magnetars are strongly magnetized neutron stars with surface magnetic fields $\sim 10^{14}-10^{15}$\,G (e.g., \citealt{Kaspi2017}). If present during the PNS phase of evolution, such strong fields have an important dynamical impact on the neutrino wind (e.g., \citealt{Prasanna2022, Prasanna2023}) with implications for heavy element nucleosynthesis \citep{Thompson2003}. Although the origin of magnetar-strength fields is debated, possible explanations include an $\alpha-\Omega$ dynamo from very rapid rotation at birth (spin periods $\lesssim 1$\,ms; \citealt{Duncan1992,Thompson1993}), strong fossil fields from the progenitor star (e.g., \citealt{Ferrario2006}), or a dynamo driven by fallback accretion (e.g., \citealt{Barrere2022}). \cite{White2022} found that even modest progenitor rotation may be sufficient to produce magnetars during core-collapse (see also \citealt{Thompson1993}). The fact that magnetars are common in the Galaxy, comprising up to $\sim 40\%$ of all neutron star births \citep{Beniamini2019}, and a lack of evidence for anomalously energetic supernova remnants surrounding magnetars \citep{Vink2006, Martin2014}, may indicate that very rapid birth periods of the order of milliseconds, are not the primary channel to form Galactic magnetars.

Several past works have explored the role that strong magnetic fields play in shaping the nucleosynthesis of proto-magnetar winds. \cite{Thompson2003} argued that a magnetar-strength dipole field is sufficient to trap a portion of the PNS wind in the equatorial closed zone of the magnetosphere; sustained energy deposition by neutrinos in this region leads to a build up of thermal pressure and the sudden ejection of high entropy material leading to a successful $r-$process (as long as the wind is neutron-rich, i.e. electron fraction $Y_{\rm e} < 0.5)$. Axisymmetric two-dimensional (2D) magnetohydrodynamic (MHD) simulations of non-rotating PNS winds by \cite{Thompson2018} confirmed the entropy enhancement and dynamical eruptions (plasmoids) from the PNS magnetosphere. \cite{Desai2023} study the dynamics of magnetized non-rotating PNS winds in three-dimensions and found entropy enhancement in plasmoids due to a combination of neutrino heating and magnetic dissipation. \cite{Vlasov2014, Vlasov2017} explore the prospects for nucleosynthesis in magnetar winds including the effects of rotation, but assume a static magnetosphere which precludes the study of plasmoid dynamics. \cite{Prasanna2022, Prasanna2023} perform a detailed study of the effects of strong magnetic fields and PNS rotation on wind dynamics, confirming high entropy regions in the dynamical plasmoid eruptions using 2D MHD simulations. They also show that magnetars born with relatively large periods $\gtrsim 100$\,ms can spindown rapidly to spin periods of the order seconds during the neutrino-driven cooling phase. \cite{Prasanna2024} extend this work to rotating magnetar winds and show how even moderately rapid rotation can influence those aspects of the wind thermodynamics that enable heavy element synthesis. In this paper, we extend the work of \cite{Prasanna2024} to study heavy element nucleosynthesis in magnetar winds by introducing tracer particles in the MHD simulations to record the thermodynamic properties of various regions of the wind, and then post-process the tracer trajectories using a nuclear reaction network to get the nucleosynthetic yields.    

The electron fraction $Y_{\rm e}$ of the PNS wind after neutrino interactions cease is a major property influencing its nucleosynthesis (e.g., \citealt{Hoffman1997}). $Y_{\rm e}$ is primarily set by the luminosities and energies of electron-type neutrinos and antineutrinos via the charged current processes (\citealt{QW1996}; but see \citealt{Metzger2008} for the effects of magnetocentrifugal slinging on $Y_{\rm e}$). Relatively recent studies have found PNS winds to be moderately neutron-rich at early times (a necessary but not sufficient condition for the $r-$process), but become proton-rich later during the cooling phase (e.g., \citealt{Roberts2012, Roberts2012_2}). The synthesis of $p-$nuclei in such proton-rich winds has received considerably attention \citep{Hoffman1996, Pruet2005, Pruet2006, Frohlich2006, Wanajo2006, Wanajo2011}, but not accounting for the effects of strong magnetic fields.

In this paper, we study heavy element nucleosynthesis in magnetar winds using two-dimensional axisymmetric fully-dynamical magnetohydrodynamic (MHD) simulations with Athena\texttt{++} \citep{Stone2020} laced with tracer particles. Tracer particles allow us to sample the thermodynamic conditions at various regions of the outflow, including the dynamical magnetosphere of proto-magnetars. We then process the thermodynamic trajectories of the tracers using the nuclear reaction network code SkyNet \citep{Lippuner2017}, and find that magnetar winds can produce a robust $r-$process extending beyond the third peak if the wind is neutron-rich. In Section \ref{section:num_setup}, we discuss the magnetar wind simulation setup including the microphysics, initial and boundary conditions, tracer particle setup in the wind simulations, and the nucleosynthesis calculation methods. In Section \ref{pre_results_section}, we briefly discuss the $r-$process mechanism, describe the structure of the PNS magnetosphere in our wind simulations, and discuss the effects of the magnetic field on the tracer paths in the simulations. In Section \ref{results_section}, we present results from the nucleosynthesis calculations. In Section \ref{discussion}, we use the results from our nucleosynthesis calculations to estimate the $r-$process element yields from magnetars. We also discuss the uncertainties in the results presented and point towards future work.    

\section{Numerical setup}
\label{section:num_setup}

\subsection{MHD equations}
We use the MHD code Athena\texttt{++} \citep{Stone2020}, which we have configured to solve the non-relativistic MHD equations:
\begin{gather}
          \frac{\partial \rho}{\partial t} + \nabla\cdot\left(\rho\bm{v}\right)=0,\label{eq:continuity}\\
          \frac{\partial \left(\rho\bm{v}\right)}{\partial t} + \nabla\cdot\left[\rho\bm{vv}+\left(P+\dfrac{B^2}{2}\right)\mathbf{I}-\bm{BB}\right]=-\rho \frac{GM_{\star}}{r^2}\boldsymbol{\hat{r}},\label{eq:momentum}\\
          \frac{\partial E}{\partial t} + \nabla\cdot\left[\left(E+\left(P+\dfrac{B^2}{2}\right)\right)\bm{v}-\bm{B}\left(\bm{B}\cdot\bm{v}\right)\right]=\dot{Q},\label{eq:energy}\\
          \frac{\partial \bm{B}}{\partial t} -\nabla\times\left(\bm{v}\times\bm{B}\right)=0,\label{eq:eulermag}
\end{gather}
where $M_{\star}$ is the mass of the PNS, $r$ is the radius from the center of the PNS, $\rho$ is the mass density of the fluid, $\bm{v}$ is the fluid velocity, $E$ is the total energy density of the fluid, $P$ is the fluid pressure, $\dot{Q}$ is the neutrino heating/cooling rate, and $\bm{B}$ is the magnetic field. We solve the MHD equations using spherical polar coordinates assuming axisymmetry. We assume that the magnetic and rotation axes of the PNS are aligned. 

\subsection{Microphysics}
\label{microphysics}
We use the Helmholtz EOS \citep{Timmes2000} to relate the pressure and density within the general equation of state (EOS) module in Athena\texttt{++} \citep{Coleman2020}\footnote{With additional modifications to enable a composition-dependent EOS.}. Since the general Helmholtz EOS is computationally expensive, we also run a few test simulations (mostly resolution tests) using an approximate analytic form of the general EOS \citep{QW1996} containing non-relativistic baryons, relativistic electrons and positrons, and photons. This is a reasonable approximation to the EOS at temperatures $T \gtrsim 0.5$\,MeV.     

We evolve the electron fraction $Y_{\rm e}$ in the MHD simulations as a function of radius and time using the rates of capture of electrons, positrons, and neutrinos on baryons \citep{QW1996}, as described in \cite{Prasanna2023}. We compute the neutrino energy deposition rate $\dot{q}$ (related to $\dot{Q}$ in equation \ref{eq:energy} as $\dot{Q}=\dot{q}\rho$) as described in \cite{Prasanna2024}. 

In this work, we include the effect on the wind dynamics and thermodynamics of the nuclear binding energy released when free nucleons are assembled to form $\alpha$ particles. To obtain the abundances of protons (p), neutrons (n), and $\alpha$ particles at a given density $\rho$, temperature $T$, and electron fraction $Y_{\rm e}$, we numerically solve the Saha equation simultaneously with baryon number conservation and charge conservation as follows \citep{Siegel2018}:
\begin{align}
    n_{\rm p}^2 n_{\rm n}^2 &= 2n_{\alpha} \left(\frac{m_{\rm b} k_{\rm B} T}{2\pi\hbar^2}\right)^{9/2} \exp \left( -Q_{\alpha}/k_{\rm B}T\right) \\
    n_{\rm b} &= n_{\rm n} + n_{\rm p} + 4n_{\alpha} \\
    n_{\rm b} Y_{\rm e} &= n_{\rm p} + 2n_{\alpha},
\end{align}
where $n_{\rm b}$, $n_{\rm n}$, $n_{\rm p}$, and $n_{\alpha}$ are number densities of baryons, neutrons, protons, and $\alpha$ particles respectively, $m_{\rm b}$ is the baryon mass, $k_{\rm B}$ is the Boltzmann constant, $\hbar$ is the reduced Planck's constant, and $Q_{\alpha} \simeq 28.3$\,MeV is the nuclear binding energy of an $\alpha$ particle. The energy difference in a given cell ($\Delta E_{\rm cell}$) due to the formation of $\alpha$ particles is accounted for using the source term in Athena\texttt{++} as follows:
\begin{equation}
\label{alpha_form_energy}
    \Delta E_{\rm cell}=Q_{\alpha}\rho N_{\rm A}\left[\nabla N_{\alpha} \cdot \bm{v}\right]dt,
\end{equation}
where $N_{\rm A}$ is Avogadro's constant, $N_{\alpha} = n_{\alpha}/n_{\rm b}$ is the number of $\alpha$ particles per baryon, $\bm{v}$ is the velocity vector, and $dt$ is the timestep.

As the free nucleons are assembled into $\alpha$ particles, the average number of nucleons per isotope $\bar{A} = \left(\sum_i X_i/A_i\right)^{-1}$ changes, where $X_i = \rho_i/\rho$ is the mass fraction of isotope $i$ and $A_i$ is the number of nucleons in isotope $i$ \citep{Timmes2000}. If we consider only protons, neutrons, and $\alpha$ particles, we have $\bar{A} = (1 - 3 N_{\alpha})^{-1}$. We pass the value of $\bar{A}$ to the Helmholtz EOS to maintain consistency. In the test simulations with the approximate analytic EOS \citep{QW1996}, $\alpha$ particle effects are neglected.  

\subsection{Reference frame, initial conditions, and boundary conditions}
\label{ICs}
We perform the simulations in an inertial (lab) frame of reference corresponding to the magnetar's center of mass. We initialize the simulations with a one dimensional non-rotating and non-magnetic (NRNM) wind model. We endow the neutron star with an initial purely dipolar magnetic field using the magnetic vector potential (see \citealt{Prasanna2022} for details). To include the effects of rotation in the initial conditions, we set the azimuthal velocity $v_{\phi}$ by assuming that angular momentum along fixed radial rays above the PNS is conserved at its value at the surface: $v_{\phi}(r)=r\Omega_{\star}\sin \theta \left(R_{\star}/r\right)^2$, where $r$ is the radius from the center of the PNS, $\theta$ is the polar angle measured from the rotation axis of the PNS, $R_{\star}$ is the PNS radius (12\,km in our simulations), and $\Omega_{\star}=2\pi/P_{\star}$ is the angular rotation frequency of the PNS. The computation grid starts from the PNS surface\footnote{In our simulations, we define the PNS surface as a sphere at a radius $R_{\star}=12$\,km from the center of the PNS.} and extends to an outer boundary radius of $10^{4}$\,km. The grid is divided into $N_r$ logarithmically spaced zones in the radial direction and $N_{\theta}$ uniformly spaced zones  in the $\theta$ direction. The boundary conditions are the same as in \cite{Prasanna2024}. We use the `first choice boundary conditions' and enforce a constant temperature at the inner boundary (see \citealt{Prasanna2024} for details).  

\subsection{Tracer particles in MHD simulations}
\label{tracers_MHD_sub}
We introduce tracer particles in the MHD simulations to track the physical and thermodynamic trajectories of the fluid parcels. Following the method used in \citet{Baronett2024}, we integrate tracer particle trajectories in Athena\texttt{++} (Chao-Chin Yang, private communication). The tracers each have position and velocity coordinates assigned. The magnetohydrodynamical properties of the fluid are interpolated to the tracers using the particle-mesh method \citep{Hockney1981}. We have made a few changes to the borrowed tracer source code such as saving the fluid data (density, temperature, electron fraction, $\bar{A}$, magnetic fields, and heating/cooling rate) of the tracers as a function of time, adjusting the tracer code for the spherical polar coordinates used in our simulations, implementing periodic release of tracers (as opposed to releasing tracers only at the start of the simulation), and implementing an MPI communication channel to release tracers from any spherical radius chosen.     

We release the tracers from a spherical radius $R_p = 15$\,km close to the surface of the PNS. We do not release the tracers from the surface of the PNS (radius of 12\,km) to avoid the discontinuities close to the surface due to insufficient radial resolution and certain approximations to the inner boundary conditions (see \citealt{Prasanna2024}). We release the tracers over time and across angular zones in such a way that all the tracers carry the same mass flux. The number of tracers released from a given angular zone is scaled to the mass flux through that zone at the time of release:
\begin{equation}
    N_p(R_p,\theta, t) = N_{\theta}\left(\dot{M}(R_p,\theta,t)/\langle \dot{M} (R_p) \rangle _{t}\right),
\end{equation}
where $N_{\theta}$ is the number of angular zones, $\dot{M}(R_p,\theta,t)$ is the mass flux through the angular zone at polar angle $\theta$, and $\langle \dot{M} (R_p) \rangle _{t}$ is the time averaged mass flux (averaged over the course of the simulation) through the spherical surface at radius $R_p$. Here, $N_{\theta}$ is just a scaling factor which ensures that there is $\sim 1$ tracer per angular zone when the mass flux through $R_p$ is the average mass flux. Mass flux through a spherical surface is given by,
\begin{equation}
    \label{mdot_eq}
        \dot{M} \left(r \right)= \oint_S r^2 \rho v_r \, {\sin \theta}d\theta d\phi.
\end{equation}

We release the tracer particles at periodic intervals of 10\,ms and record the tracer magnetohydrodynamic properties such as density, temperature, electron fraction, neutrino heating rate, and magnetic fields along the tracer path. The tracer snapshots are output every 0.5\,ms. We confirm that our results are not sensitive to these choices, and we obtain similar results if we double the time interval between tracer releases or halve the cadence of tracer snapshots.

We employ the nuclear reaction network SkyNet \citep{Lippuner2017} to process the tracer trajectories recorded from the MHD simulations as described above. We use the self-heating function in SkyNet, which evolves the network taking into account the binding energy release from nuclear reactions \citep{Lippuner2017}. We start from a composition at nuclear statistical equilibrium (NSE) because the tracers are released from a region of high temperature and density ($\rho \gtrsim 10^9$\,g cm$^{-3}$, $T \gtrsim 2$\,MeV). Since the self-heating function only takes into account the density profile of the tracer, it has no knowledge of entropy enhancements along the tracer path due to neutrino heating or magnetic dissipation in the MHD simulations (see Sections \ref{mag_structure} and \ref{tracer_paths_sub}). We input this information as an external heating term in SkyNet as follows:
\begin{equation}
    \dot{q}_{\rm ext} = T_{\rm avg} \Delta S / \Delta t - \dot{q}_{\alpha}
\end{equation}
where $\Delta S$ is the change in entropy along the tracer path in a time $\Delta t$ (mostly 0.5\,ms in our simulations, which is the time interval between tracer snapshots), $T_{\rm avg}$ is the average temperature between the two tracer snapshots, and $\dot{q}_{\alpha}$ is the rate of energy release due to formation of $\alpha$ particles (see equation \ref{alpha_form_energy}). We subtract the alpha-formation energy in SkyNet to avoid double counting as the self-heating function in SkyNet itself takes into account energy released from nuclear reactions. We find that this method produces good agreement between the entropy along the tracer path in the MHD simulation and the entropy evolved within SkyNet. 

As mentioned above, we initialize the composition in SkyNet assuming nuclear statistical equilibrium (NSE) using the values of density $\rho_i$ and temperature $T_i$ at the point of tracer release, and for various values of electron fraction $Y_{{\rm e},i}$ (see Tables \ref{table1} and \ref{table2}). Once the initial composition is set, subsequent changes to $Y_{\rm e}$ are assumed to occur only due to nuclear reactions within the network. Thus, we ignore changes to $Y_{\rm e}$ that occur along the tracer path in the MHD simulations, e.g. due to alpha particle formation. This is a good approximation because $Y_{\rm e}$ typically freezes out due to weak interactions relatively close to the PNS surface (see \citealt{Thompson2001,Prasanna2022, Prasanna2023}), changing by less than a few percent along the tracer path. The wind electron fraction is mostly determined by the neutrino luminosities and energies \citep{QW1996}, which we set by hand. Our choice to vary by hand the initial $Y_{{\rm e},i}$ used in our network calculations without self-consistently running new MHD simulations is reasonable because modest changes to $Y_{{\rm e},i} \sim 0.4-0.5$ are expected to only moderately impact the wind dynamics and evolution as a result of small changes to the wind heating rate and pressure (on the other hand, changes to $Y_{\rm e, i}$ have a much larger impact on the nucleosynthesis). 

The tracers typically require $\sim 0.5$\,s to reach the outer boundary of the grid. At later times, we extrapolate the density profile as $\rho \propto t^{-a}$, with $a=3$. This scaling, close to homologous expansion, is justified by an extrapolation of the tracer density evolution near the outer boundary of our NRNM simulations. We confirm that our nucleosynthesis results are not sensitive to changes in $a$ in the range $1-4$. We terminate the SkyNet calculations at a network time of $t_{\rm end} = 10^{9}$\,s, though our results for the final abundances are not sensitive to this choice because most of the remaining nuclei are stable by this time.

The above extrapolation of density does not take into consideration the possibility of interaction between the expanding magnetar wind and reverse shocks, such as those that can occur when the supernova shock passes through the Si-O interface \citep{Janka1995, Thompson2001}. Such an interaction can increase the entropy and expansion timescale of the wind \citep{Thompson2001}. Such an interaction may have potential implications for nucleosynthesis, such as increasing the yield of the elements in the third $r-$process peak \citep{Thompson2001}. We intend to study the effects of such shocks on nucleosynthesis in a future work.  

\section{Magnetosphere structure and Tracer paths}
\label{pre_results_section}
\subsection{Path to the $r-$process}
\label{rprocess_sub}

Before presenting our numerical results, we briefly review the physical stages leading to an $r-$process in PNS winds. Neutron star wind material is initially composed of free nucleons, electrons, positrons, neutrinos, and photons close to the PNS surface. As the wind material expands away from the PNS and cools, free nucleons are assembled into $\alpha$ particles. Around a temperature region $T\sim 0.5$\,MeV ($\sim 5\times 10^{9}$\,K), nearly all the protons are used up to form $\alpha$ particles in neutron-rich winds. Subsequently, these alpha particles undergo a neutron-aided triple-alpha-like process to form $^{12}$C and then an $\alpha$ process occurs to burn the $\alpha$ particles to heavy nuclei (\citealt{Hoffman1997}; see also Section 4 of \citealt{Meyer1998}). Once the charged particle reactions freeze at $\sim 0.2$\,MeV, the $\alpha$ process ends. The heavy nuclei produced from the $\alpha$ process stage become the seed nuclei for the subsequent rapid neutron capture stage.   

In order to synthesize the heaviest $r-$process elements, extending up to the third mass peak at $\rm A \sim 195$, the ratio of neutrons to seed nuclei must be sufficiently high.  For relatively high $Y_{\rm e} \gtrsim 0.4$ relevant to PNS winds, this requires suppressing the formation of seed nuclei at the $^{12}$C formation stage.  A rough criterion on the wind properties just after $\alpha$ particle formation ($T \approx 0.5$ MeV) to achieve a third-peak $r-$process can be written (\citealt{Hoffman1997}):
\begin{equation}
    \label{zeta_eqn}
    \zeta = \frac{S^{3}}{1.28\,Y_{\rm e}^{3}t_{\rm exp}} \ge \zeta_{\rm crit} = 8\times 10^{9}\, \rm (k_{B} \ baryon^{-1})^{3} \ s^{-1}
\end{equation} 
where $S$ is the entropy and $t_{\rm exp}$ is the outflow expansion timescale.  Thus, for a given $Y_{\rm e}$, sufficiently high entropy and/or rapid expansion can enable a heavy $r-$process.  The sensitive dependence of $\zeta \propto S^{3}$ implies that small changes to the wind entropy when $\zeta \sim \zeta_{\rm crit}$ can lead to large changes in the synthesized abundances of heavy nuclei with $\rm A \ge 195$.  This ``threshold'' effect will influence some of our results described below.

\subsection{Structure of the PNS magnetosphere}
\label{mag_structure}
Figure 1 of \citet{Prasanna2024} shows the structure of the PNS magnetosphere for various polar magnetic field strengths and neutrino luminosities. The magnetic field structure is found to be monopolar when the wind pressure gradient is much larger than the magnetic tension in the equatorial region. On the other hand, when the magnetic tension is comparable to the wind pressure gradient, the wind material becomes trapped in the closed zone of the magnetic field. The trapped material eventually gets ejected when the wind pressure gradient aided by neutrino heating overcomes the magnetic tension. As the heating of the trapped matter by neutrinos also increases the entropy relative to that of a free wind, this leads to quasi-periodic ejections of high-entropy plasma \citep{Thompson2003, Thompson2018, Prasanna2024}. \cite{Prasanna2024} used diagnostic parameters such as the entropy and expansion rate through the seed formation region (Eq.~\ref{zeta_eqn}) to show that the high-entropy ejections in their simulations are potentially favorable for heavy element nucleosynthesis. 

\begin{figure*}
\centering{}
\includegraphics[width=\textwidth]{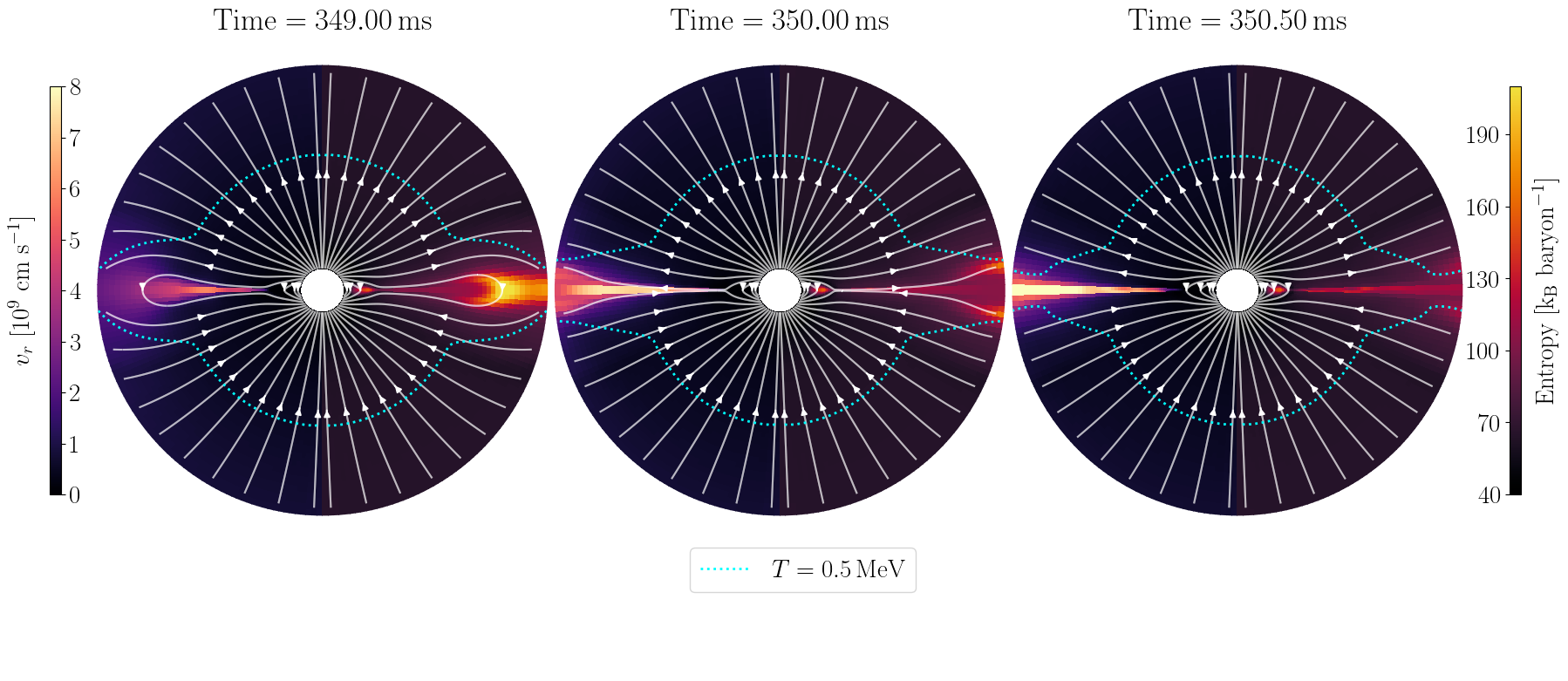}
\vspace{-14mm}
\caption{Three snapshots (left to right) from our fiducial simulation (at a polar magnetic field $B_0=4\times 10^{15}$\,G for a non-rotating PNS) showing 2D maps of radial velocity $v_r$ (left half of each panel) and entropy (right half of each panel). We show the structure of the magnetic field during magnetic reconnection and subsequent plasmoid eruption. The outer boundary shown here is at a radius of 100\,km (though the full simulation domain extends to $10^{4}$\,km). The white lines are the magnetic field lines and the central white circle is the PNS with a radius of 12\,km. We show the $T=0.5$\,MeV surface (dotted cyan lines) because this is the temperature around which the $\alpha-$process occurs to burn the $\alpha$ particles into heavier seed nuclei for the subsequent $r-$process, and the properties of the wind around this temperature are critical for nucleosynthesis. See the figures in \cite{Prasanna2024} for the effects of various values of magnetic field and neutrino luminosity on the structure of the magnetosphere.} 
\label{plasmoids_sequence}
\end{figure*}

Figure \ref{plasmoids_sequence} shows the structure of the magnetic field close to the PNS surface during a plasmoid eruption, as well as the $T=0.5$\,MeV surface critical for nucleosynthesis (Section \ref{rprocess_sub}). A current sheet (which is not resolved in our simulations) separates the upper and lower hemispheres of the magnetosphere, where magnetic field lines of opposite polarity cross outside the closed zone. Such a structure is a common feature of pulsar and neutron star magnetospheres (e.g., \citealt{Bucciantini2006, Uzdensky2014, Cerutti2015, Hu2022, Hakobyan2023}). As shown in Figure \ref{plasmoids_sequence}, heating from the neutrinos blows a bubble of high entropy material as the wind pressure gradient increases to catch up with the magnetic tension in the equatorial region of the PNS. Ultimately, when the wind pressure gradient exceeds the magnetic tension, the magnetic field reconfigures itself by reconnecting, which leads to ejection of the trapped material which now has acquired high entropy \citep{Thompson2003}. The resistivity of the current sheet in our MHD simulations is numerical. As we shall describe below, the entropy of the ejected material is increased not only by neutrino heating, but also by the energy released during magnetic reconnection. 

\subsection{Tracer paths}
\label{tracer_paths_sub}

\begin{figure}
\centering{}
\includegraphics[width=0.82\linewidth]{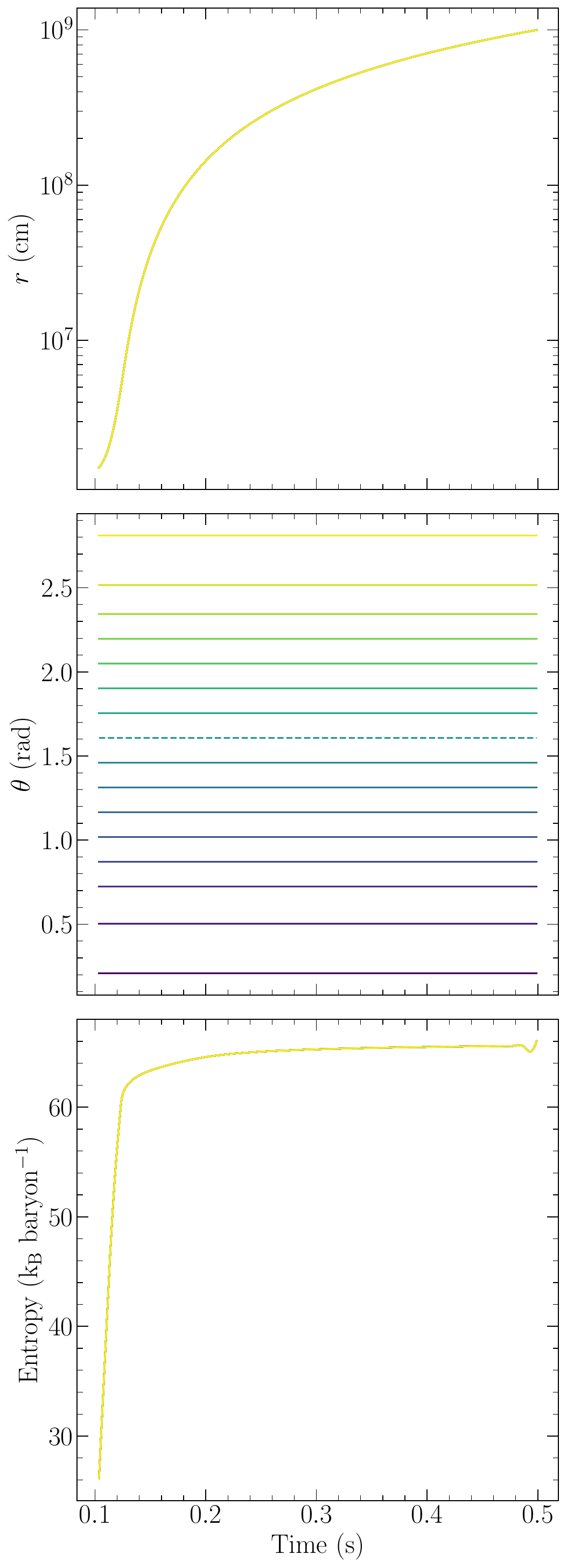}
\caption{Radial location (top panel), polar angle (middle panel), and entropy (bottom panel) of a few tracers from the wind simulation of a non-rotating and non-magnetic (NRNM) PNS. We obtain a spherically symmetric steady state wind in this case. The radial coordinate and entropy profiles of the tracers as a function of time are the same for all tracers released at the same instant of time from various polar angles. The trajectory closest to the PNS equator in the plot is shown with a dashed line.} 
\label{tracer_paths_B0}
\end{figure}

\begin{figure}
\centering{}
\includegraphics[width=0.82\linewidth]{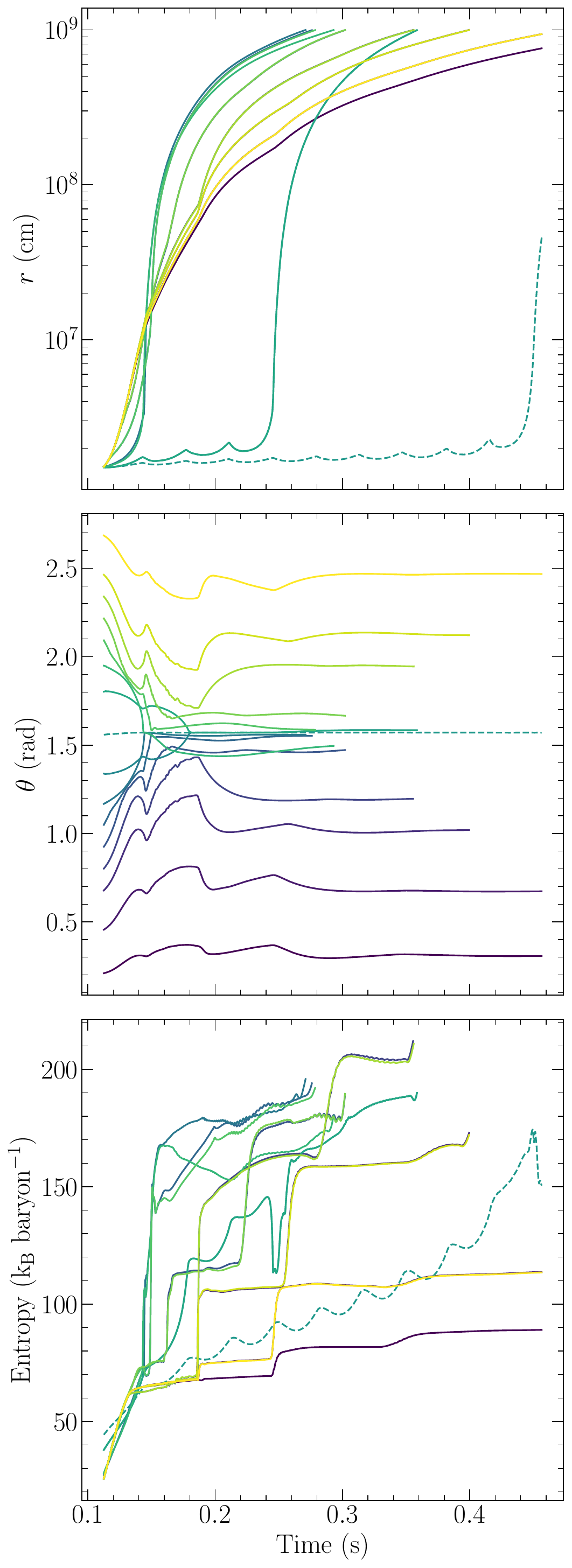}
\caption{Radial location (top panel), polar angle (middle panel), and entropy (bottom panel) of a few tracers from the wind simulation of a non-rotating PNS at a polar magnetic field strength $B_0=4\times 10^{15}$\,G. Based on the polar angle of the initial tracer release, various physical and thermodynamic tracer paths are possible as shown. Tracers released close to the PNS equator are trapped for a significant amount of time before being released into the expanding wind. The trajectory closest to the PNS equator in the plot is shown with a dashed line.} 
\label{tracer_paths}
\end{figure}

Figure \ref{tracer_paths_B0} shows the radial location, polar angle, and entropy of a few tracers released from different polar angles, as a function of time, from the non-rotating non-magnetic (NRNM) wind simulation. As expected, absent strong magnetic fields, we obtain a steady state spherically symmetric wind. Tracers released from different polar angles at the same time experience similar dynamic and thermodynamic properties as shown. This contrasts with the tracer trajectories of the magnetized non-rotating PNS wind with $B_0=4\times 10^{15}$\,G shown in Figure \ref{tracer_paths}. The dynamical and thermodynamic paths of the tracers in the magnetized case are strongly dependent on the initial polar angle of release. Tracers initially released near the poles, along which the magnetic field lines are consistently open, move outwards freely. Their physical paths are only slightly affected by the plasmoid ejections in the equatorial region. As was described in Section \ref{mag_structure}, the outflowing material is trapped in the equatorial region, as evidenced by the more erratic green tracer paths in Figure \ref{tracer_paths}. Some tracers oscillate back and forth in radius for a significant amount of time before being ejected from the magnetic closed zone. Such tracers gain entropy from neutrino heating as shown in the bottom panel of Figure \ref{tracer_paths}.      

We find that the tracers gain entropy not only from neutrino heating, but also from energy dissipation associated with magnetic reconnection and  mixing of fluid parcels. This is evident in the bottom panel of Figure \ref{tracer_paths}, which shows sharp spikes in entropy. The spikes that occur within a few hundred kilometers from the PNS surface correspond to magnetic energy dissipation from reconnection. The entropy spikes that occur at thousands of kilometers from the PNS surface correspond to mixing of different fluid parcels as the plasmoids expand along the equatorial region. Although we are certain that energy dissipation due to magnetic reconnection increases the entropy of wind material, we do not know how much of this entropy increase is physical, due to lack of a physical model for the current sheet in our MHD simulations. For this reason, we present results from nucleosynthesis calculations considering the total entropy (entropy gain from all sources) from the simulations at face value, and separately for calculations in which only the entropy gain from neutrino heating is considered. Although our calculations that include neutrino heating alone likely underestimate the formation of heavy $r-$process elements, we present these results as a conservative lower bound on $r-$process yields of proto-magnetars.

Another potential complication with our nucleosynthesis calculations is mixing of fluid parcels. Tracers may not accurately represent the thermodynamic conditions of the fluid if significant mixing occurs prior to or during key stages of the nucleosynthesis, such as the seed formation phase or neutron-capture phase as the material expands and cools. Figure \ref{tracer_paths} shows entropy spikes that arise from mixing occur near the outer computational boundary; however, these occur well after the formation of $\alpha$ particles and hence are not expected to significantly affect the seed production.  Although the nucleosynthesis results presented in this paper neglect mixing, the effects of mixing on the nucleosynthesis warrant a more detailed study in future work.

\section{Results}
\label{results_section}
\begin{figure*}
\centering{}
\includegraphics[width=\textwidth]{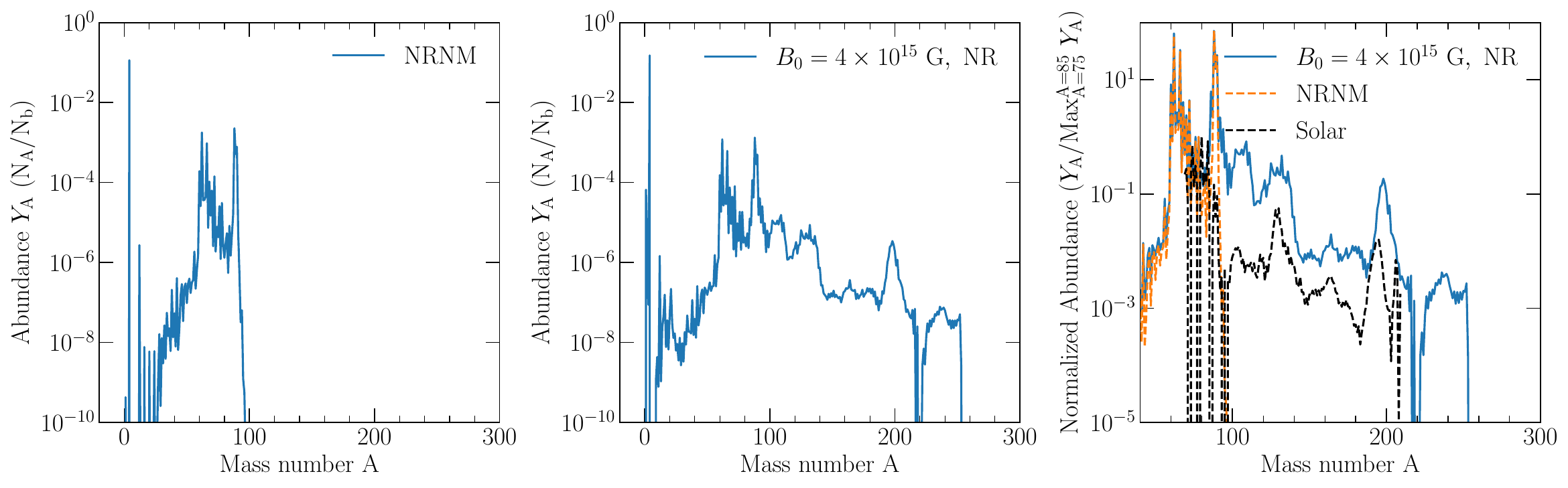}
\caption{The left panel shows average abundance per baryon for a non-rotating, non-magnetic (NRNM) PNS while the middle panel shows average abundance per baryon for a non-rotating PNS at a polar magnetic field $B_0=4\times 10^{15}$\,G. The abundance distributions shown are averages of those obtained by all the tracers. The right panel compares the abundance patterns (normalized to the first $r-$process peak) in these two cases with the Solar $r-$process abundance pattern \citep{Arnould2007}. Note the difference in the y-axis range of the right panel compared to the left and the middle panels. The value of electron fraction used to set the initial composition in the SkyNet nucleosynthesis calculations is $Y_{\rm e}\sim 0.47$ (see Tables \ref{table1} and \ref{table2}). The EOS used in the MHD simulation is the HEOS and the resolution is $256\times 128$ for all the calculations shown in this plot.} 
\label{skynet_B_comp}
\end{figure*}

We now present results from our nucleosynthesis calculations. Figure \ref{skynet_B_comp} shows the final abundances per baryon $Y_{\rm A} = \rm{N_A/N_b}$ in the left and middle panels, where $\rm N_A$ is the number of ions with mass number A (not to be confused with the Avogadro's constant $N_{\rm A}$ in Eq. \ref{alpha_form_energy}) and $\rm N_b$ is the number of baryons, averaged over the yields of all the tracers. First consider the non-rotating, non-magnetic (NRNM) wind shown in the left panel. Consistent with earlier works (e.g., \citealt{QW1996}), we see that the NRNM wind fails to produce a robust $r-$process extending up to the third peak. The middle panel of Figure \ref{skynet_B_comp} shows results for a magnetized but non-rotating PNS wind with $B_0=4\times 10^{15}$\,G (the fiducial model), which yields a robust $r-$process extending beyond the third peak. The right panel compares the abundance patterns of these two models with the Solar $r-$process abundance pattern, all normalized to the abundance of the first $r-$process peak. 

Tables \ref{table1} and \ref{table2} presents results of our nucleosynthesis calculations for different values of PNS magnetic field $B_0$, spin period $P_{\star}$, and neutrino luminosity. Most models assume $L_{\bar{\nu}_{\rm e}}=8\times 10^{51}$\,ergs s$^{-1}$, corresponding to roughly $1-2$\,s into the cooling evolution of the PNS \citep{Pons1999, Vartanyan2023}; the mean $\bar{\nu}_{\rm e}$ neutrino energy, as well as the luminosities and energies of the other neutrino flavors, follow those described in \cite{Prasanna2024} (their Section 3.3). The mass loss rate of the wind, $\dot{M}$, decreases rapidly in time as the PNS cools \citep{Thompson2001, Prasanna2022}, so we are most interested in the prospects for nucleosynthesis when $\dot{M}$ is largest during the earliest phases in the PNS cooling evolution. We also present results for lower neutrino luminosities $L_{\bar{\nu}_{\rm e}}=4\times 10^{51}$\,ergs s$^{-1}$ and $L_{\bar{\nu}_{\rm e}}=3\times 10^{51}$\,ergs s$^{-1}$, corresponding to $\sim 5-6$\,s after the explosion, and $L_{\bar{\nu}_{\rm e}}=2\times 10^{51}$\,ergs s$^{-1}$ and $L_{\bar{\nu}_{\rm e}}=8.9\times 10^{50}$\,ergs s$^{-1}$, which occur $\sim 10-15$\,s after the explosion, to explore the implications for nucleosynthesis later in the cooling epoch. The neutrino luminosities and energies of all the neutrino species for the lower luminosity calculations are based on the PNS cooling model of \cite{Vartanyan2023}, as described in Section 3.5 of \cite{Prasanna2024}.  

Tables \ref{table1} and \ref{table2} also presents results for different assumptions about the wind electron fraction $Y_{\rm e}$ (see Section \ref{tracers_MHD_sub}). In some models we take the initial $Y_{\rm e}$  in our network calculations to be the time average of the electron fraction along the tracer path, while in other cases we vary the initial $Y_{\rm e}$ by hand in the range $0.4-0.5$. As expected, for otherwise similar dynamic and thermodynamic properties, the yields of heavy $r-$process elements are found to be greater for more neutron-rich winds (lower $Y_{\rm e,i}$). 

As described in Section \ref{tracer_paths_sub}, the entropy gained by the outgoing fluid can in general arise from a combination of sources, including neutrino heating, magnetic energy dissipation, and fluid mixing. Since we do not have a physical model for the current sheet and hence the amount of entropy enhancement from magnetic dissipation is uncertain, we present results separately for two different assumptions regarding the entropy input to our network calculations.  In one case we use the total entropy of the MHD simulations taken at face value ($\dot{q}_{\rm ext} \ {\rm mode}=H_{\rm all}$ in Tables \ref{table1} and \ref{table2}), while in the other case we use only the entropy gained from neutrino heating ($\dot{q}_{\rm ext} \ {\rm mode}=H_{\nu}$ in Tables \ref{table1} and \ref{table2}).  

\begin{deluxetable*}{cccccccccccc}
\tablecolumns{12}
\label{table1}
\tablewidth{0pt}
\tablecaption{Summary of SkyNet nucleosynthesis calculations. To avoid clutter and confusion, we use quotes (") to represent a table entry which is same as the corresponding entry in the previous row.}
\tablehead{
    \colhead{\tablenotemark{\rm \scriptsize a}$B_0$} & \colhead{\tablenotemark{\rm \scriptsize b}$P_{\star}$} & \colhead{\tablenotemark{\rm \scriptsize c}$L_{\bar{\nu}_{\rm e}}$} & \colhead{\tablenotemark{\rm \scriptsize d}EOS} & \colhead{\tablenotemark{\rm \scriptsize e}$Y_{\rm e}$} & \colhead{\tablenotemark{\rm \scriptsize f}Resolution} & \colhead{\tablenotemark{\rm \scriptsize g}$\dot{q}_{\rm ext}$ mode} & \colhead{\tablenotemark{\rm \scriptsize h}$N_{\rm tr,\, avg}$} & \colhead{\tablenotemark{\rm \scriptsize i}$\dot{M}_{\rm tr}$} &  \colhead{\tablenotemark{\rm \scriptsize j}$\dot{M}_{\rm A>190}$} & \colhead{\tablenotemark{\rm \scriptsize k}$\dot{M}_{\rm A>130}$} & \colhead{\tablenotemark{\rm \scriptsize l}$\dot{M}_{\rm A>79}$}\\   
    ($10^{15}$\,G) & (ms) & ($\rm 10^{51} \ ergs \ s^{-1}$)  & & &  & & & $(\rm M_{\odot} \ s^{-1})$ & $(\rm M_{\odot} \ s^{-1})$ & $(\rm M_{\odot} \ s^{-1})$ & $(\rm M_{\odot} \ s^{-1})$  
    }

    \startdata
    0.0 & \tablenotemark{\rm \scriptsize m}NR & 8  & HEOS & \tablenotemark{\rm \scriptsize n}avg  & LR & $H_{\rm all}$ & 186 & $2\times10^{-6}$ & $0.0$ & $0.0$ & $1.2\times 10^{-4}$   \\ \\
    0.5 & 20 & 0.89  & HEOS & 0.44  & LR & $H_{\rm all}$ & 290 & $3\times10^{-8}$ & $7.3\times 10^{-9}$ & $2.2\times 10^{-8}$ & $4.2\times 10^{-6}$   \\
    " & " & "  & " & 0.42  & " & " & " & " & $1.1\times 10^{-8}$ & $3.8\times 10^{-8}$ & $4.7\times 10^{-6}$   \\
    " & " & "  & " & 0.40  & " & " & " & " & $2.1\times 10^{-8}$ & $8.4\times 10^{-8}$ & $5.0\times 10^{-6}$   \\ \\
    1 & 20 & 3  & HEOS & 0.42  & LR & $H_{\rm all}$ & 97 & $1\times10^{-6}$ & $9.1\times 10^{-10}$ & $3.0\times 10^{-8}$ & $6.7\times 10^{-5}$   \\ 
    " & " & "  & " & 0.40  & " & " & " & " & $7.7\times 10^{-9}$ & $1.4\times 10^{-7}$ & $7.2\times 10^{-5}$   \\ 
    " & 8 & "  & " & "  & " & " & 112 & " & $2.1\times 10^{-7}$ & $1.1\times 10^{-6}$ & $7.9\times 10^{-5}$   \\ 
    " & 20 & 2  & " & 0.42  & " & " & 215 & $1.9\times10^{-7}$ & $1.4\times 10^{-7}$ & $4.1\times 10^{-7}$ & $2.4\times 10^{-5}$   \\ 
    " & " & "  & " & 0.40  & " & " & " & " & $2.2\times 10^{-7}$ & $7.9\times 10^{-7}$ & $2.6\times 10^{-5}$   \\ \\
    3 & 200 & 8  & HEOS &  avg  & LR & $H_{\rm all}$ & 145 & $2\times 10^{-6}$ & $5.2\times 10^{-7}$ & $2.3\times 10^{-6}$ & $7.5\times 10^{-5}$   \\
    " & " & "  & " &  0.44  & " & " & " & " & $2.0\times 10^{-6}$ &  $7.4\times 10^{-6}$ & $1.6\times 10^{-4}$   \\
    " & " & "  & " &  "  & " & $H_{\nu}$ & " & " & $1.6\times 10^{-8}$ & $8.5\times 10^{-7}$ & $1.6\times 10^{-4}$   \\
    " & " & "  & " &  0.42  & " & " & " & " & $1.8\times 10^{-7}$ & $2.4\times 10^{-6}$ & $1.8\times 10^{-4}$   \\
    " & " & 4  & " &  0.44  & " & $H_{\rm all}$ & 235 & $7.8\times 10^{-7}$ & $1.9\times 10^{-6}$ & $5.5\times 10^{-6}$ & $9.3\times 10^{-5}$   \\
    " & " & "  & " &  "  & " & $H_{\nu}$ & " & " & $8.6\times 10^{-8}$ & $1.2\times 10^{-6}$ & $9.4\times 10^{-5}$   \\
    " & " & "  & " &  0.42  & " & " & " & " & $3.8\times 10^{-7}$ & $2.7\times 10^{-6}$ & $1.0\times10^{-4}$   \\
    " & 20 & 8  & " &  avg  & " & $H_{\rm all}$ & 152 & $2\times 10^{-6}$ & $6.9\times 10^{-7}$ & $2.5\times 10^{-6}$ & $1.0\times10^{-4}$   \\
    " & " & "  & " &  0.44  & " & " & " & " & $2.3\times 10^{-6}$ & $7.7\times 10^{-6}$ & $1.6\times 10^{-4}$   \\
\enddata
\tablenotetext{\rm \scriptsize a}{ Polar magnetic field of the PNS.}
\tablenotetext{\rm \scriptsize b}{ Spin period of the PNS.}
\tablenotetext{\rm \scriptsize c}{ Luminosity of the electron type anti-neutrino.}
\tablenotetext{\rm \scriptsize d}{ The equation of state (EOS) used for the MHD simulation. QW refers to the approximate analytic form of the general EOS \citep{QW1996} while the HEOS refers to the Helmholtz EOS \citep{Timmes2000}. All the SkyNet nucleosynthesis calculations use the Helmholtz EOS (see \citealt{Lippuner2017}) irrespective of the EOS used in the MHD simulation.}
\tablenotetext{\rm \scriptsize e}{ Value of the electron fraction $Y_{\rm e}$ used in the SkyNet nucleosynthesis calculations to set the initial composition (see Section \ref{tracers_MHD_sub}).}
\tablenotetext{\rm \scriptsize f}{ The MHD simulations are run on a grid with $N_r$ logarithmically spaced radial zones and $N_{\theta}$ evenly spaced angular zones. $N_r\times N_{\theta}$ for the entries marked with LR, MR, and HR are respectively $256\times 128$, $512\times 256$, and $1024\times 512$.}
\tablenotetext{\rm \scriptsize g}{ $\dot{q}_{\rm ext}$ is the external heating function we feed into SkyNet to account for the entropy gain along the tracer path. Mode $H_{\rm all}$ corresponds to $\dot{q}_{\rm ext}$ calculated from total entropy while mode $H_{\nu}$ corresponds to $\dot{q}_{\rm ext}$ calculated from entropy gain from neutrinos only.}
\tablenotetext{\rm \scriptsize h}{ Average number of tracers per release.}
\tablenotetext{\rm \scriptsize i}{ Mass flux carried by each tracer.}
\tablenotetext{\rm \scriptsize j}{ Average mass flux of elements of with $\rm A>190$.}
\tablenotetext{\rm \scriptsize k}{ Average mass flux of elements of with $\rm A>130$.}
\tablenotetext{\rm \scriptsize l}{ Average mass flux of elements of with $\rm A>79$.}
\tablenotetext{\rm \scriptsize m}{ \  A non-rotating (NR) PNS.}
\tablenotetext{\rm \scriptsize n}{ SkyNet calculation in which $Y_{\rm e}$ of the initial composition is set by the average electron fraction along the tracer path.}
\end{deluxetable*}

\begin{deluxetable*}{cccccccccccc}
\tablecolumns{12}
\tablewidth{0pt}
\label{table2}
\tablecaption{Continuation of Table \ref{table1} (see footnotes of Table \ref{table1} for the notation). To avoid clutter and confusion, we use quotes (") to represent a table entry which is same as the corresponding entry in the previous row.}
\tablehead{
    \colhead{\tablenotemark{\rm \scriptsize a}$B_0$} & \colhead{\tablenotemark{\rm \scriptsize b}$P_{\star}$} & \colhead{\tablenotemark{\rm \scriptsize c}$L_{\bar{\nu}_{\rm e}}$} & \colhead{\tablenotemark{\rm \scriptsize d}EOS} & \colhead{\tablenotemark{\rm \scriptsize e}$Y_{\rm e}$} & \colhead{\tablenotemark{\rm \scriptsize f}Resolution} & \colhead{\tablenotemark{\rm \scriptsize g}$\dot{q}_{\rm ext}$ mode} & \colhead{\tablenotemark{\rm \scriptsize h}$N_{\rm tr,\, avg}$} & \colhead{\tablenotemark{\rm \scriptsize i}$\dot{M}_{\rm tr}$} &  \colhead{\tablenotemark{\rm \scriptsize j}$\dot{M}_{\rm A>190}$} & \colhead{\tablenotemark{\rm \scriptsize k}$\dot{M}_{\rm A>130}$} & \colhead{\tablenotemark{\rm \scriptsize l}$\dot{M}_{\rm A>79}$}\\   
    ($10^{15}$\,G) & (ms) & ($\rm 10^{51} \ ergs \ s^{-1}$)  & & &  & & & $(\rm M_{\odot} \ s^{-1})$ & $(\rm M_{\odot} \ s^{-1})$ & $(\rm M_{\odot} \ s^{-1})$ & $(\rm M_{\odot} \ s^{-1})$
    }

    \startdata
    3 & 20 & 8  & QW &  avg  & LR & $H_{\rm all}$ & 142 & $2\times 10^{-6}$ & $1.1\times 10^{-6}$ & $4.1\times 10^{-6}$ & $1.2\times 10^{-4}$   \\
    " & " & "  & " &  0.42  & " & " & " & " & $3.3\times 10^{-6}$ & $1.2\times 10^{-5}$ & $1.6\times 10^{-4}$   \\
    " & " & "  & " &  avg  & MR & " & 296 & $1\times10^{-6}$ & $5.4\times 10^{-7}$ & $2.3\times 10^{-6}$ & $1.3\times 10^{-4}$   \\
    " & " & "  & " &  0.42  & " & " & " & " & $1.9\times 10^{-6}$ & $8.7\times 10^{-6}$ & $1.7\times 10^{-4}$   \\
    " & " & "  & " &  avg  & HR & " & 628 & $5\times 10^{-7}$ & $4.8\times 10^{-7}$ & $2.5\times 10^{-6}$ & $1.4\times 10^{-4}$   \\
    " & " & "  & " &  0.44  & " & " & " & " & $8.9\times 10^{-7}$ & $4.4\times 10^{-6}$ & $1.6\times 10^{-4}$   \\
    " & " & "  & " &  0.42  & " & " & " & " & $1.9\times 10^{-6}$ & $8.7\times 10^{-6}$ & $1.8\times 10^{-4}$   \\
    " & 10 & "  & HEOS &  avg  & LR & " & 170 & $2\times 10^{-6}$ & $8.7\times 10^{-7}$ & $2.8\times 10^{-6}$ & $1.2\times 10^{-4}$   \\
    " & " & "  & " &  0.44  & " & " & " & " & $2.7\times 10^{-6}$ & $8.8\times 10^{-6}$ & $1.6\times 10^{-4}$   \\ 
    " & " & "  & " &  0.42  & " & $H_{\nu}$ & " & " & $2.2\times 10^{-7}$ & $3.6\times 10^{-6}$ & $1.9\times 10^{-4}$   \\ \\
    4 & NR & 8  & HEOS &  avg  & LR & $H_{\rm all}$ & 126 & $2\times 10^{-6}$ & $1.1\times10^{-6}$ &  $2.6\times10^{-6}$ & $6.0\times10^{-5}$   \\
    " & " & "  & " &  "  & " & $H_{\nu}$ & " & " & $3.4\times10^{-8}$ & $2.4\times10^{-7}$ & $7.9\times10^{-5}$   \\
    " & " & "  & " &  0.42  & " & " & " & " & $1.2\times10^{-6}$ &  $5.0\times10^{-6}$ &  $1.5\times10^{-4}$   \\
    " & " & "  & " &  0.49  & " & $H_{\rm all}$ & " & " & $4.0\times10^{-7}$ & $8.1\times10^{-7}$ & $5.8\times10^{-6}$    \\
    " & " & "  & " &  0.44  & " & " & " & " & $2.8\times10^{-6}$ & $8.0\times10^{-6}$ & $1.3\times10^{-4}$   \\
    " & " & "  & " &  0.42  & " & " & " & " & $4.2\times10^{-6}$ & $1.5\times10^{-5}$ & $1.5\times10^{-4}$   \\ \\
    4 & NR & 8  & QW &  avg  & LR & $H_{\rm all}$ & 113 & $2\times 10^{-6}$ & $1.6 \times 10^{-6}$ & $4.6 \times 10^{-6}$ & $1.0\times10^{-4}$   \\
    " & " & "  & " &  "  & " & $H_{\nu}$ & " & " & $1.2 \times 10^{-7}$ & $9.9 \times 10^{-7}$ & $1.0\times10^{-4}$   \\
    " & " & "  & " &  0.42  & " & " & " & " & $9.5 \times 10^{-7}$ & $4.1 \times 10^{-6}$ & $1.3 \times 10^{-4}$   \\
    " & " & "  & " &  avg  & MR & $H_{\rm all}$ & 256 & $1\times10^{-6}$ & $1.0\times10^{-6}$ & $5.0 \times 10^{-6}$ & $1.1 \times 10^{-4}$   \\
    " & " & "  & " &  "  & HR & " & 559 & $5\times 10^{-7}$ & $4.3\times 10^{-7}$ & $2.0\times 10^{-6}$ & $1.2\times 10^{-4}$   \\
    " & " & "  & " &  "  & " & $H_{\nu}$ & " & " & 0 & $1.0\times10^{-9}$ & $1.2\times 10^{-4}$   \\
    " & " & "  & " &  0.42  & " & " & " & " & $3.0 \times 10^{-8}$ & $8.7 \times 10^{-7}$ & $1.6 \times 10^{-4}$   \\ \\
    6 & NR & 6.5  & QW &  avg  & LR & $H_{\rm all}$ & 126 & $1\times10^{-6}$ & $1.1\times 10^{-6}$ & $2.2\times 10^{-6}$ & $5.4\times 10^{-5}$   \\
    " & " & "  & " &  "  & " & $H_{\nu}$ & " & " & $1.5\times 10^{-7}$ & $7.0\times 10^{-7}$ & $5.5\times 10^{-7}$   \\
    " & " & "  & " &  "  & HR & $H_{\rm all}$ & 579 & $2.5\times 10^{-7}$ & $6.3\times 10^{-7}$ & $1.5\times 10^{-6}$ & $6.7\times 10^{-5}$   \\
    " & " & "  & " &  "  & " & $H_{\nu}$ & " & " & $1.7\times 10^{-8}$ & $2.3\times 10^{-7}$ & $6.6\times 10^{-5}$   \\
\enddata
\end{deluxetable*}

\begin{figure*}
\centering{}
\includegraphics[width=\textwidth]{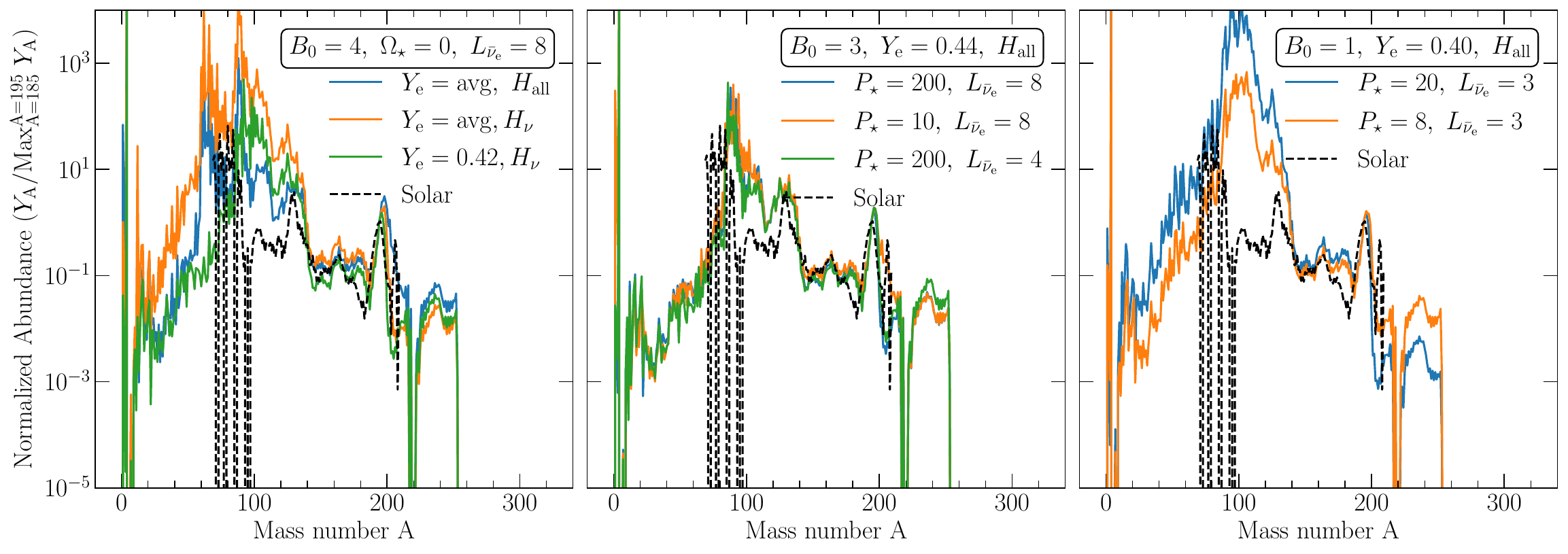}
\caption{Normalized abundance pattern for different values of PNS polar magnetic field $B_0$ (in $10^{15}$\,G), spin period $P_{\star}$ (in millisecond; $\Omega_{\star}=2\pi/P_{\star}=0$ indicates a non-rotating PNS), neutrino luminosity $L_{\bar{\nu}_{\rm e}}$ (in $10^{51}$\,ergs s$^{-1}$), electron fraction $Y_{\rm e}$ used to the set the initial composition in the nucleosynthesis calculations, and external heating function input to SkyNet $\dot{q}_{\rm ext}$ ($H_{\rm all}$ indicates calculations which use the total entropy while $H_{\nu}$ indicates calculations which consider heating from neutrinos alone; see Tables \ref{table1} and \ref{table2} and Section \ref{results_section}). The EOS used in the MHD simulation is the HEOS and the resolution is $256\times 128$ for all the calculations shown in this plot.} 
\label{abund_all_plots}
\end{figure*}

\begin{figure*}
\centering{}
\includegraphics[width=\textwidth]{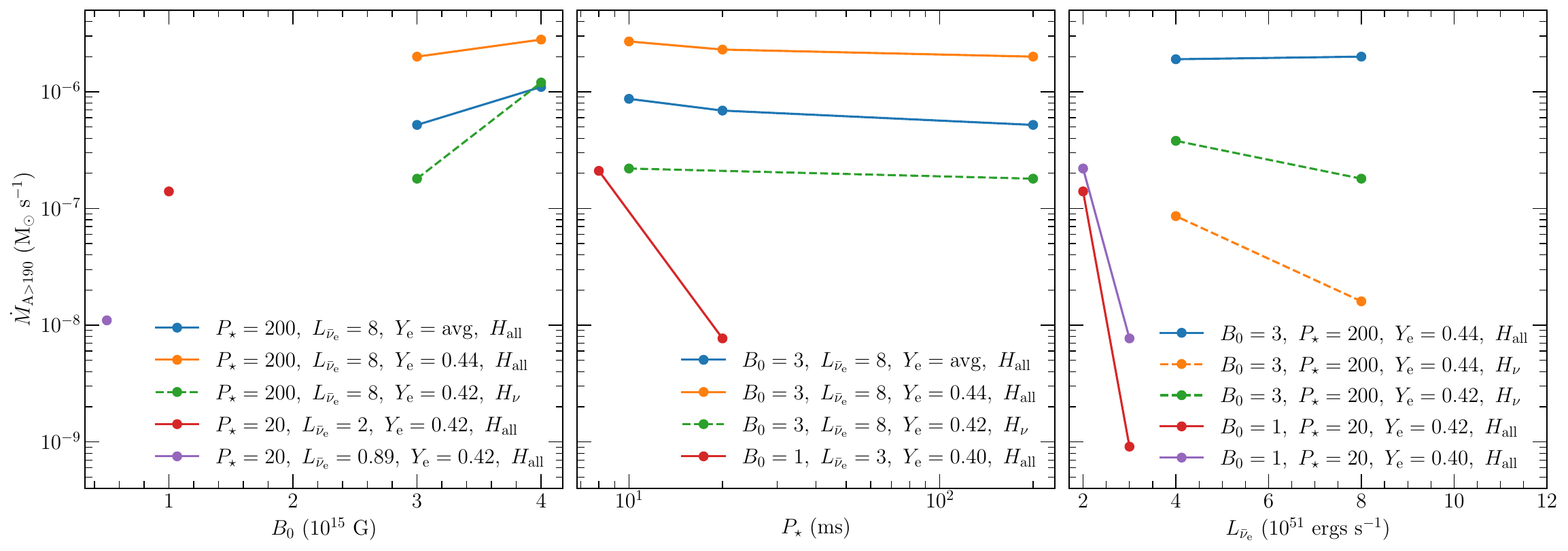}
\caption{Mass flux of elements with mass number $\rm A>190$ as a function of polar magnetic field (left panel), PNS spin period (middle panel), and neutrino luminosity (right panel). $P_{\star}$ is the PNS spin period in millisecond, $B_0$ is the PNS magnetic field in units of $10^{15}$\,G, $L_{\bar{\nu}_{\rm e}}$ is the neutrino luminosity in units of $10^{51}$\,ergs s$^{-1}$, and $Y_{\rm e}$ is the electron fraction of the initial composition in the SkyNet nucleosynthesis calculation. The solid lines labeled with $H_{\rm all}$ represent nucleosynthesis calculations that consider total entropy while the dashed lines labeled with $H_{\nu}$ are the calculations that consider entropy from neutrino heating only (see Tables \ref{table1} and \ref{table2}). The EOS used in the MHD simulation is the HEOS and the resolution is $256\times 128$ for all the calculations shown in this plot.} 
\label{abund_A190}
\end{figure*}

Figure \ref{abund_all_plots} shows abundance patterns for the calculations presented in Tables \ref{table1} and \ref{table2}, normalized to the third $r-$process peak at $\rm A=195$.  All models exhibit a robust $r-$process up to and beyond the third peak. Although the solar $r-$pattern is reproduced reasonably well close to the third peak, we find a large overproduction of lighter elements with $\rm A \lesssim 120$, which varies inversely with the mass yield of elements with $\rm A>190$. This overproduction was also found in previous works on $r-$process nucleosynthesis in PNS winds (e.g., \citealt{Woosley1994, Hoffman1996_2, Pruet2006}). In our simulations, the overproduction of light $\rm A \lesssim 120$ elements chiefly arises from matter with lower entropy $\lesssim 100$\,$\rm k_B \ baryon^{-1}$ released from high latitudes away from the equatorial region, i.e. those parts of the wind which most resemble an ordinary NRNM wind. This overproduction is likely sensitive to the electron fraction of the wind, which we remind is put in by hand in our calculations. 

Figure \ref{abund_A190} shows the mass flux of third-peak elements with $\rm A>190$ as a function of the PNS polar magnetic field $B_0$, spin period $P_{\star}$, and neutrino luminosity $L_{\bar{\nu}_{\rm e}}$. As expected, the relative abundances of third peak elements increases with the magnetic field, because the entropy achieved by matter ejected from the equatorial closed magnetospheric zone increases with magnetic field. Confirming a conclusion of \cite{Prasanna2024}, we find that more rapid PNS rotation enhances the yield of third-peak elements (middle panel in Figure \ref{abund_A190}), as particularly evident for the $B_0=10^{15}$\,G model (see Table \ref{table1}). A PNS rotating at 8\,ms yields $\sim 30$ times more $\rm A>190$ elements versus the $P_{\star} =$ 20\,ms case. As we describe in the next section, this large sensitivity follows from a ``threshold'' effect related to the entropy being close to the critical value for third-peak nucleosynthesis (see also Section \ref{rprocess_sub}).

\subsection{Effects of resolution}
\label{resolution_sub}
\begin{figure}
\centering{}
\includegraphics[width=0.82\linewidth]{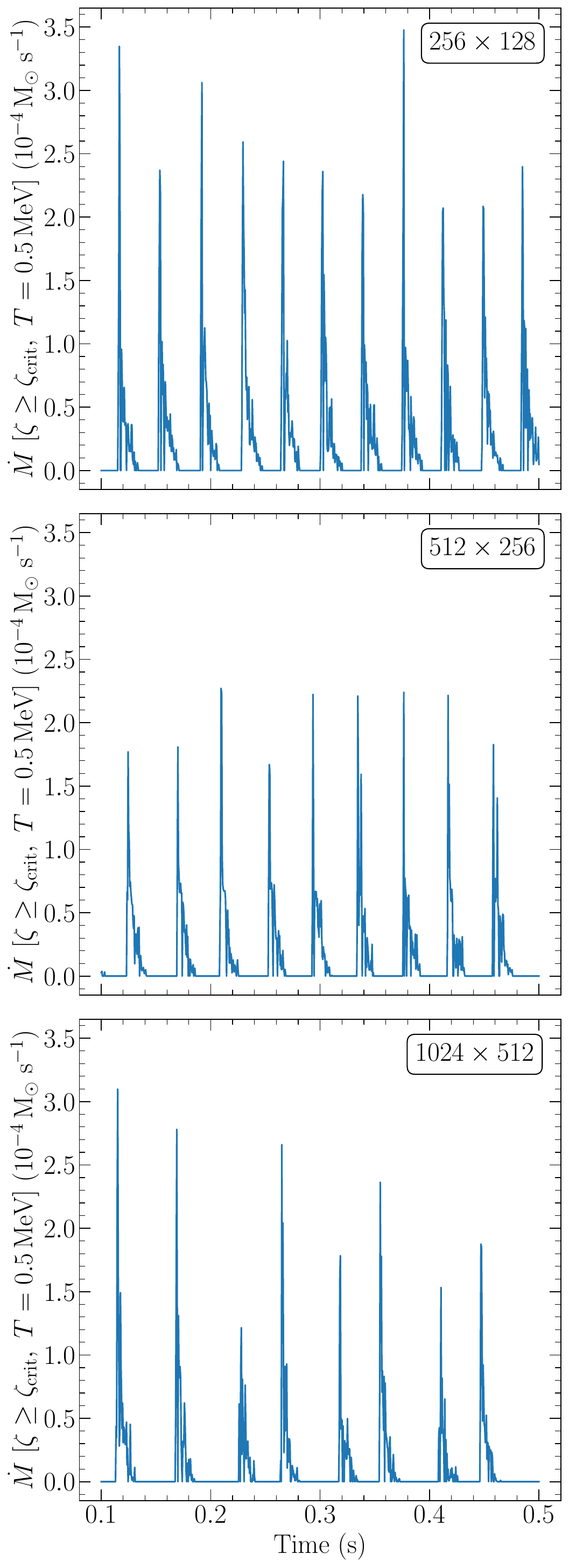}
\caption{Mass flux (estimated using the hydrodynamic quantities) through the $T=0.5$\,MeV surface that satisfies the Hoffman criterion (Eq.~\ref{zeta_eqn}) for production of the third $r-$process peak \citep{Hoffman1997} as a function of resolution (indicated as $N_r\times N_{\theta}$ in the top-right corner of each panel).} 
\label{mdot_zeta}
\end{figure}

\begin{figure}
\centering{}
\includegraphics[width=\linewidth]{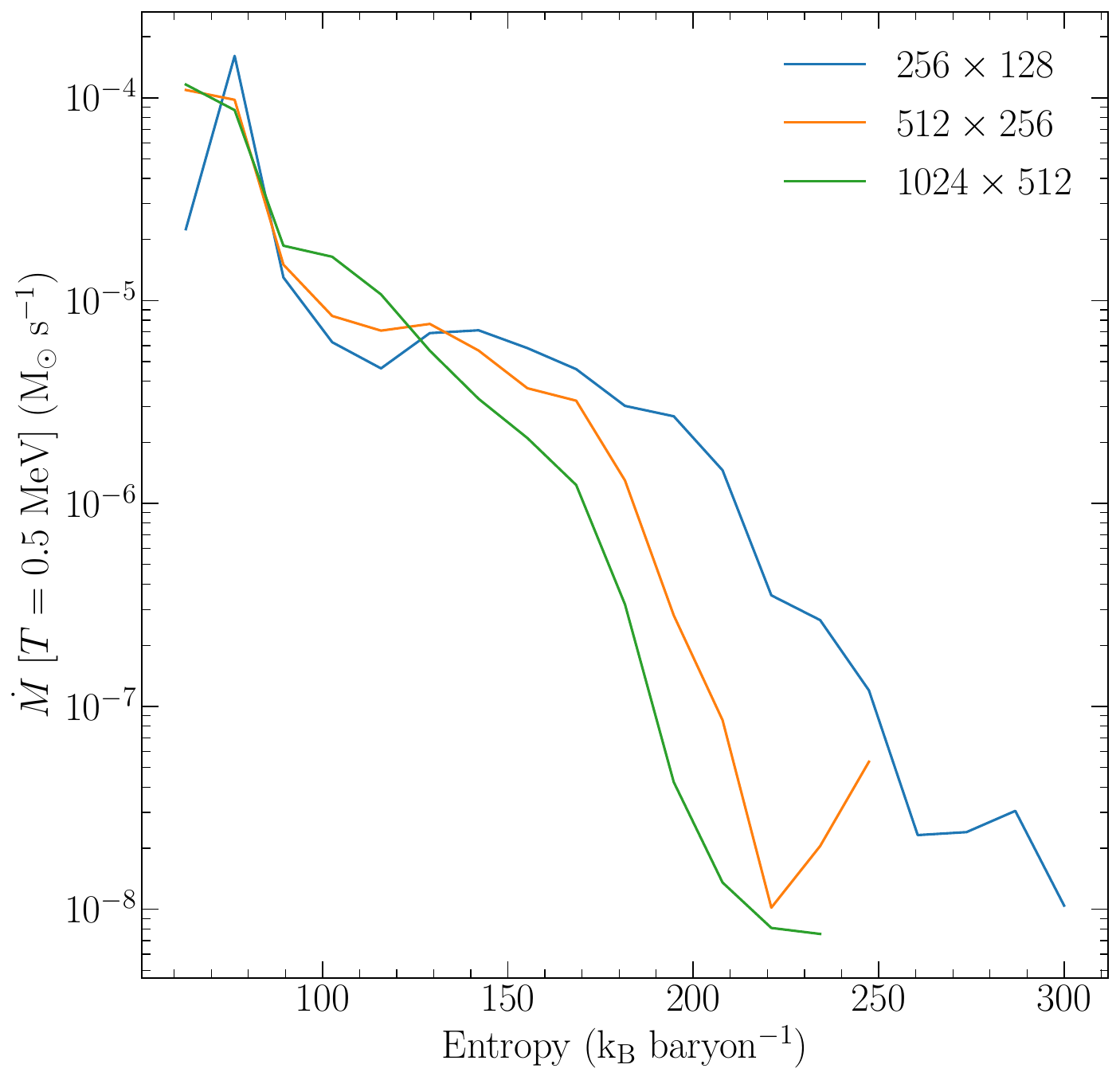}
\caption{Time-averaged mass flux (estimated using the hydrodynamic quantities) through the $T=0.5$\,MeV surface as a function of entropy for various resolutions (indicated as $N_r\times N_{\theta}$ in the top-right corner).} 
\label{entr_bins_hydro}
\end{figure}

Our MHD simulations reveal a highly dynamic PNS magnetosphere (Section \ref{mag_structure}), in which periodic ejection of high-entropy plasmoids occur from the equatorial closed zone region. However, the details of the plasmoid ejection process and its implications for nucleosynthesis are found to depend on the resolution of our simulations. To demonstrate this, Figure \ref{mdot_zeta} shows the mass flux of wind material through the $T=0.5$\,MeV surface (estimated using the hydrodynamic quantities) that satisfies the analytic ``Hoffman'' criterion for the production of third peak $r-$process elements (Eq.~\ref{zeta_eqn}).  Figure \ref{mdot_zeta} shows that the time interval separating peaks in the mass flux satisfying the Hoffman criterion is resolution dependent. Figure \ref{entr_bins_hydro} likewise shows the entropy distribution of the time-averaged mass flux through the $T=0.5$\,MeV surface, illustrating that lower resolution simulations yield higher entropy wind material, thereby explaining their greater propensity to satisfy Eq.~\ref{zeta_eqn} and generate larger quantities of third-peak elements.      


\begin{figure*}
\centering{}
\includegraphics[width=\textwidth]{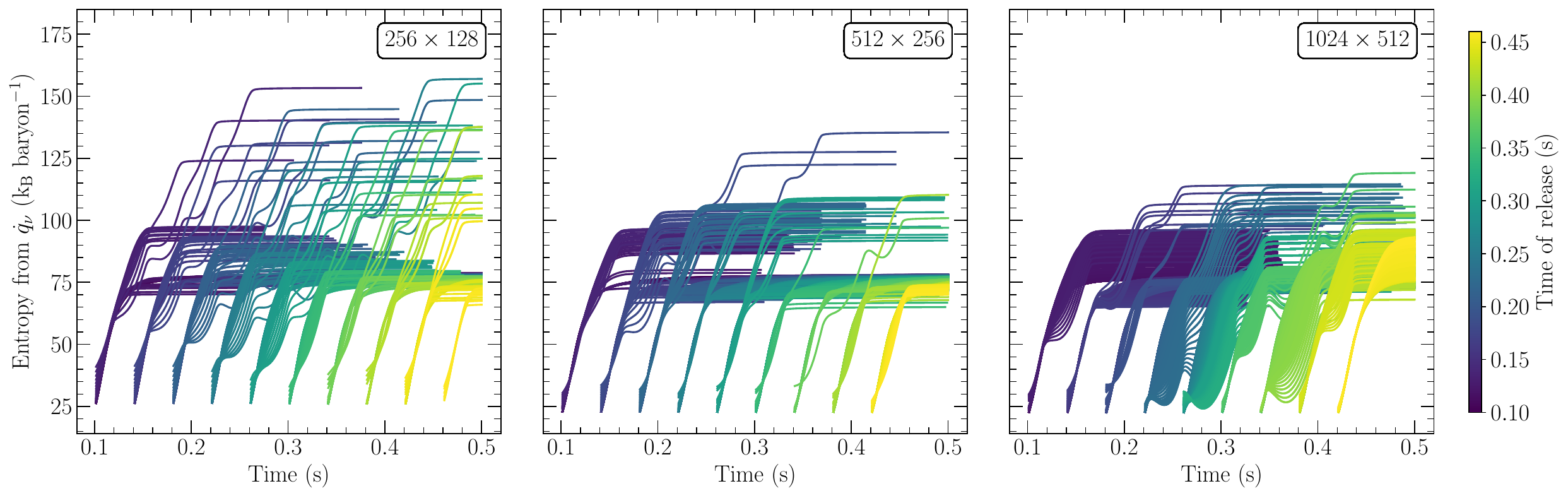}
\caption{Entropy along the tracer path for different resolutions (indicated as $N_r\times N_{\theta}$ at the top right of each panel) at a PNS polar magnetic field $B_0=4\times 10^{15}$\,G for a non-rotating PNS. Entropy shown here is from neutrino heating only (mode $H_{\nu}$ in Tables \ref{table1} and \ref{table2}).} 
\label{rescomp_B4e15_QdotOnly}
\end{figure*}

Figure \ref{rescomp_B4e15_QdotOnly} shows the profiles of tracer entropy arising from neutrino heating alone. The peak entropy achieved by the tracers is seen to decrease with higher resolution, although the differences are not drastic (amounting to $\lesssim 20-30$\,${\rm k_B \ baryon^{-1}}$). As long as the wind entropy greatly exceeds the threshold for third-peak $r-$process (for a given $Y_{\rm e}$ value), these differences do not have a large qualitative effect on the final abundance pattern. On the other hand, if the entropy of the tracers is close to the threshold value for reaching the third $r-$process peak (Eq.~\ref{zeta_eqn}), the effects become more dramatic. One such example is the high resolution (HR) calculation at $B_0=4\times 10^{15}$\,G with the QW EOS and $\dot{q}_{\rm ext} \  {\rm mode} = H_{\nu}$ (considering entropy from neutrino heating only). For the model with $Y_{\rm e} \sim 0.46$ (marked with ``avg'' in Table \ref{table2}), the yield of $\rm A>190$ elements is $\dot{M}_{\rm A>190}=0$, versus $\dot{M}_{\rm A>190}=1.2\times 10^{-7}$\,M$_{\odot} \ \rm s^{-1}$ for a similar calculation at lower resolution (LR; see Table \ref{table2}). This difference can be attributed to the lower tracer entropy achieved by the HR calculation compared to the LR case, as shown in Figure \ref{rescomp_B4e15_QdotOnly}.           
To verify that this sensitivity to resolution is indeed an entropy ``threshold'' effect, we ran additional calculations for a higher magnetic field $B_0=6\times 10^{15}$\,G but smaller neutrino luminosity $L_{\bar{\nu}_e}=6.5 \times 10^{51}$\,ergs $\rm s^{-1}$. When considering the entropy gain from neutrino heating alone ($H_{\nu}$; see Table \ref{table2}) in the HR calculation, the yield of $\rm A>190$ elements is non-zero, unlike the otherwise similar case for $B_0=4\times 10^{15}$\,G. In other words, the marginal increase in tracer entropy due to the slightly stronger magnetic field now enables a successful third $r-$process peak.

As a further test, we ran additional wind calculations for smaller $Y_{\rm e}$, keeping everything else fixed. In those simulations using the QW EOS, the average electron fraction along the tracer path is $\sim 0.45-0.46$, versus an average $\sim 0.47$ in simulations using the Helmholtz EOS \citep{Timmes2000}.\footnote{The slightly higher average electron fraction for the Helmholtz EOS is due to the formation of $\alpha$ particles, which are neglected in the QW EOS.} The HR calculation for $B_0=4\times 10^{15}$\,G and $\dot{q}_{\rm ext} {\rm \ mode} \ H_{\nu}$ exhibits a non-zero yield of $\rm A>190$ elements for $Y_{\rm e}=0.42$, in contrast to the calculation with average electron fraction along the tracer path (see Table \ref{table2}). As another example, we ran calculations with $Y_{\rm e}=0.42$ at $B_0=3\times 10^{15}$\,G and $P_{\star}=20$\,ms using the QW EOS. The difference in the third-peak element yields with resolution is smaller when compared to those using $Y_{\rm e}$ averaged along the tracer path (see Table \ref{table2}). Taken together, we conclude that the dependence of the yield of third-peak $r-$process elements on resolution mainly arises from the resolution dependence of the entropy when the latter is close to the critical threshold value required for a heavy $r-$process.

Although drastic differences in the nucleosynthesis yield of elements with mass number $\rm A>190$ are seen as a function of resolution in a few calculations with $\dot{q}_{\rm ext} \ {\rm mode} =H_{\nu}$ (considering neutrino heating alone; see Table \ref{table2}) mainly due to the threshold effect described above, our calculations with $\dot{q}_{\rm ext} \ {\rm mode} =H_{\rm all}$ (considering total entropy) yield mostly similar results with the difference in the heavy $r-$process yield ($\dot{M}_{\rm A>190}$) being a factor of $\lesssim 2-3$. For the models with $B_0=3\times 10^{15}$\,G, $P_{\star}=20$\,ms, and $\dot{q}_{\rm ext} \ {\rm mode} = H_{\rm all}$, the value of $\dot{M}_{\rm A>190}$ is found to converge well with increasing resolution, although this is not as true for $B_0\geq 4\times 10^{15}$\,G (Table \ref{table2}). However, in both the cases, $\dot{M}_{\rm A>130}$ shows much less variability across resolutions compared to the values of $\dot{M}_{\rm A>190}$, while the values of $\dot{M}_{\rm A>79}$ match very well across resolutions (see Tables \ref{table1} and \ref{table2}). 

Using 2D axisymmetric relativistic MHD simulations of rotating magnetized neutron stars, \cite{Bucciantini2006} found that the formation and growth of plasmoids in the equatorial belt of the magnetosphere depends on the radial magnetic field $B_r$, velocity along the polar direction $v_{\theta}$, and the rate of change of $B_r$ with the polar angle. We speculate that some of the resolution dependence seen in our simulations arises from these quantities changing close to the current sheet, and that convergence should occur for further increase in resolution. However, this cannot be tested due to the computational cost of higher resolution simulations. Future work will explore the properties of the current sheet in greater detail. For purposes of this paper, we must settle for a reasonable convergence in our results.

\subsection{Effects of electron fraction $Y_{\rm e}$}
\begin{figure}
\centering{}
\includegraphics[width=\linewidth]{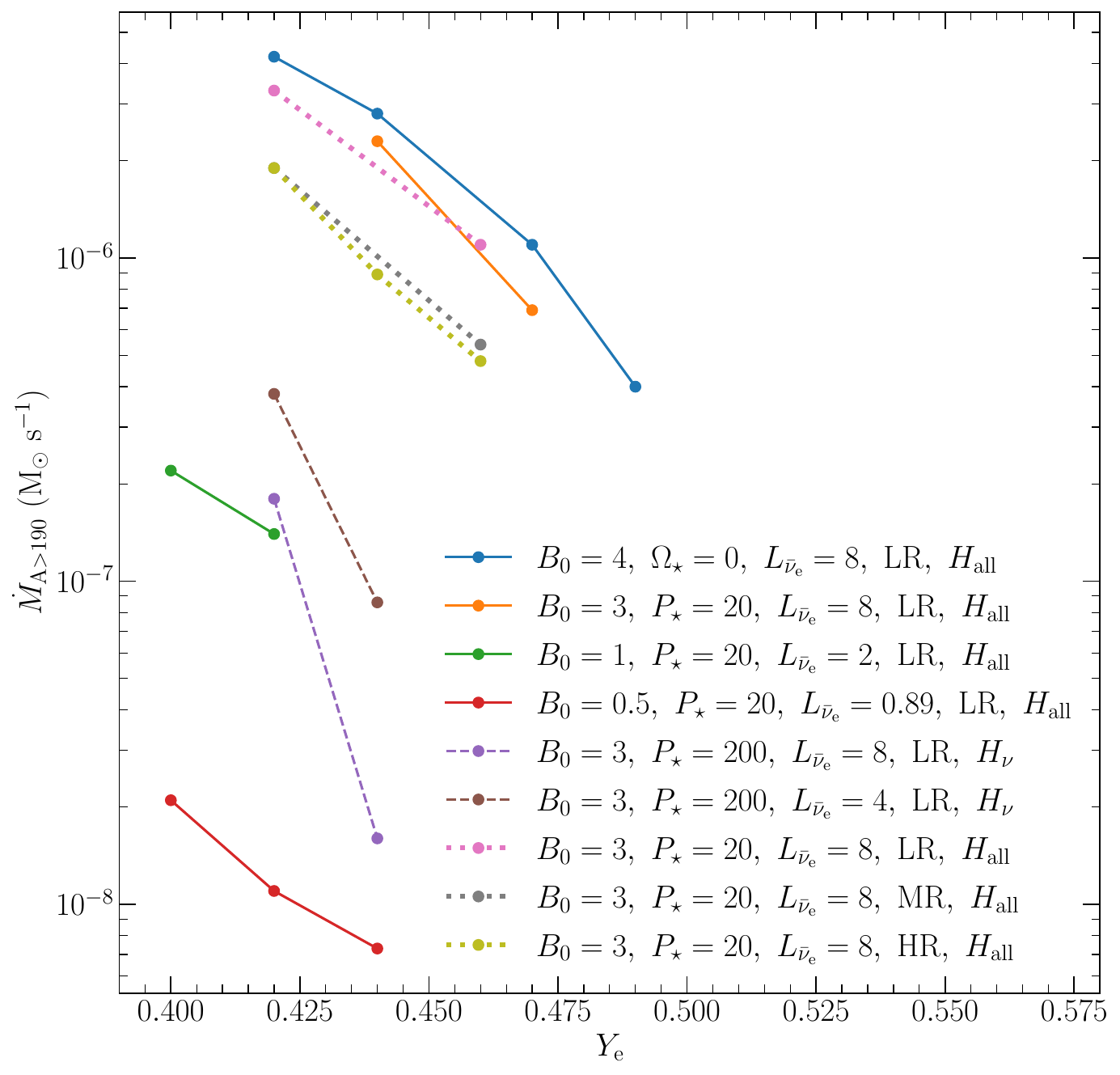}
\caption{Mass flux of elements with mass number $\rm A>190$ as a function of electron fraction $Y_{\rm e}$ of the initial composition in the SkyNet nucleosynthesis calculations. The dotted lines are the MHD simulations that use the QW equation of state \citep{QW1996}, while all the other lines are the MHD simulations that use the Helmholtz equation of state (see Tables \ref{table1} and \ref{table2}). The dashed lines are the nucleosynthesis calculations which consider entropy gain of the wind material (in the MHD simulations) from neutrino heating only (mode $H_{\nu}$, see Tables \ref{table1} and \ref{table2}), while all the other lines consider entropy gain from all sources (mode $H_{\rm all}$, see Tables \ref{table1} and \ref{table2}). The line labels in order are the PNS magnetic field $B_0$ (in units of $10^{15}$\,G), spin period $P_{\star}$ (in millisecond; $\Omega_{\star}=2\pi/P_{\star}=0$ indicates a non-rotating PNS), neutrino luminosity $L_{\bar{\nu}_{\rm e}}$ (in units of $10^{51}$\,ergs s$^{-1}$), resolution (indicated as LR, MR, or HR; see Tables \ref{table1} and \ref{table2}), and external heating input to the nucleosynthesis calculations.} 
\label{md_ye}
\end{figure}

The electron fraction $Y_{\rm e}$ is one of the critical properties of the outflow affecting heavy element nucleosynthesis. Previous studies have found that lower $Y_{\rm e}$ at a given entropy and expansion timescale provide more favorable conditions to produce the third $r-$process peak in neutron-rich winds (except around $Y_{\rm e} \sim 0.48-0.5$, e.g., \citealt{Hoffman1997}). We find similar trends in our SkyNet nucleosynthesis calculations. Figure \ref{md_ye} shows the yield of heavy $\rm A>190$ elements as a function of the electron fraction $Y_{\rm e}$ used to set the initial composition in the nucleosynthesis calculations for various values of magnetic field, PNS spin period, neutrino luminosity, equation of state, and resolution. From this plot and Tables \ref{table1} and \ref{table2}, it is evident that lower value of $Y_{\rm e}$ while everything else is held fixed provides more favorable conditions to produce the third $r-$process peak.  

\begin{figure}
\centering{}
\includegraphics[width=\linewidth]{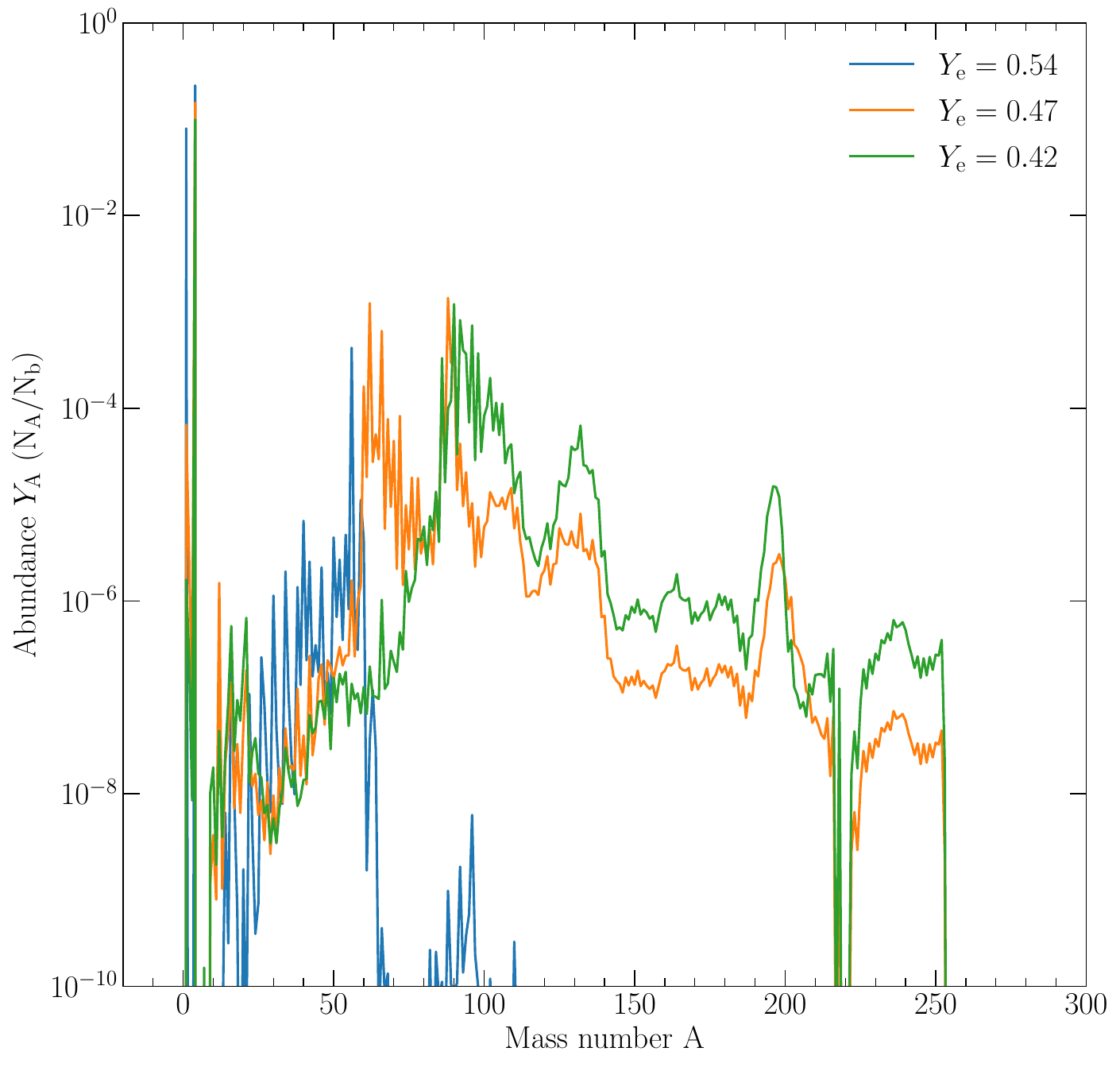}
\caption{Average abundance per baryon for various values of electron fraction $Y_{\rm e}$ used in the nucleosynthesis calculations to set the initial composition. The MHD simulation corresponding to these nucleosynthesis calculations has been run with the Helmholtz EOS for a non-rotating PNS with a polar magnetic field $B_0=4\times 10^{15}$\,G (at a resolution of $256\times 128$).} 
\label{skynet_abund_Ye}
\end{figure}

Figure \ref{skynet_abund_Ye} shows the final abundance pattern for different values of electron fraction $Y_{\rm e}$ setting the initial composition, for a non-rotating PNS with a polar magnetic field $B_0=4\times 10^{15}$\,G. As expected, for neutron-rich winds ($Y_{\rm e} < 0.5$), lower values of $Y_{\rm e}$ increase the abundance of elements close to the third $r-$process peak. On the other hand, in the case of a proton-rich wind ($Y_{\rm e} = 0.54$) the distribution of elements synthesized is drastically different, even though the properties of the PNS and the MHD wind are otherwise the same in all the three cases. 

Note that in the proton-rich calculation, we do not incorporate a potential flux of neutrons that may originate from antineutrino captures on protons \citep{Pruet2006}. Such a flux of neutrons can help bridge long waiting points along the nucleosynthesis path. Nucleosynthesis in proton-rich winds will be explored in greater detail in in a future work.  

\begin{figure}
\centering{}
\includegraphics[width=\linewidth]{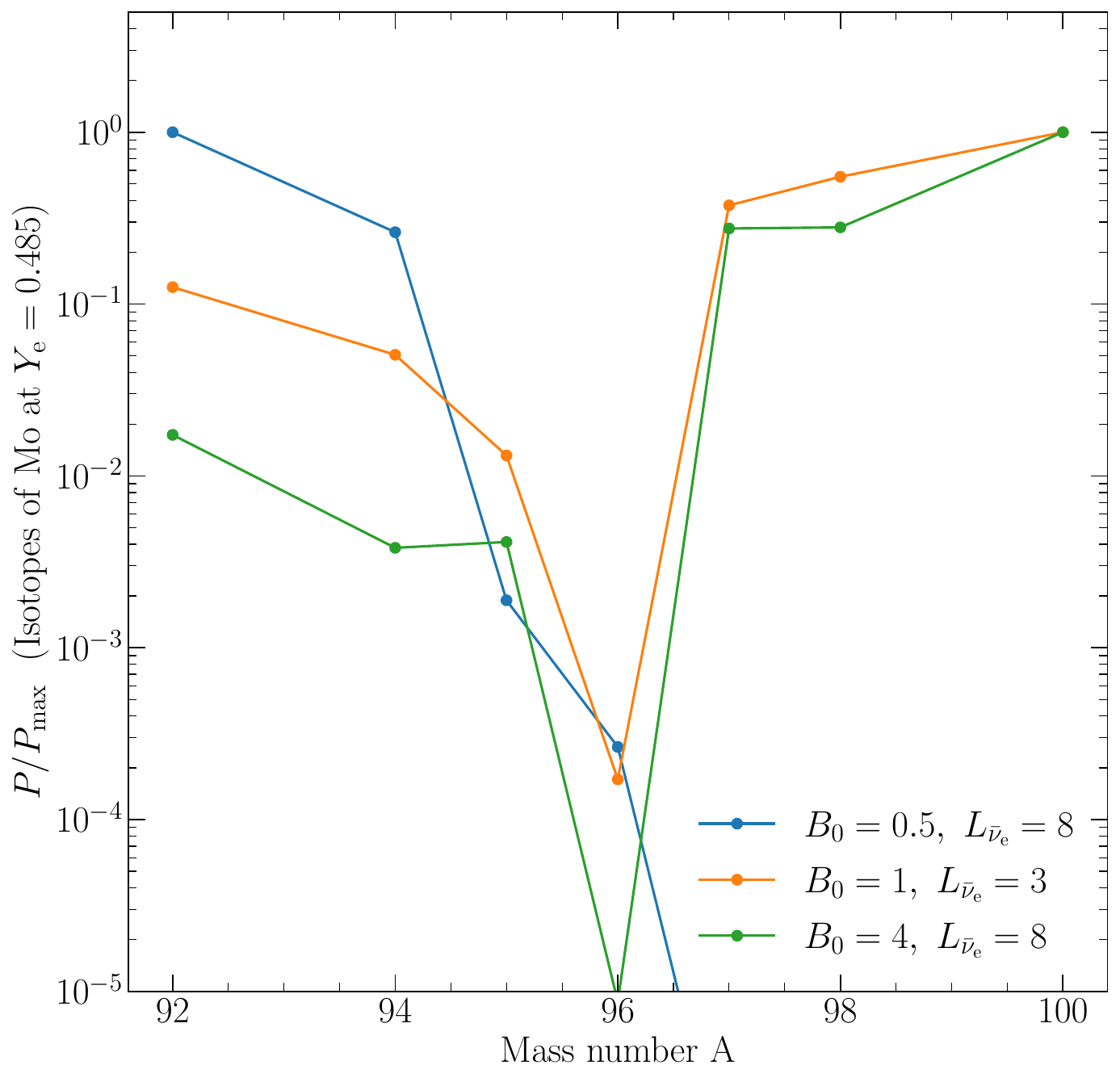}
\caption{Production factor (Eq. \ref{prod_factor}) of various isotopes of Molybdenum (Mo) normalized to the maximum production factor of the isotopes of Mo. The line labels are PNS polar magnetic field $B_0$ in units of $10^{15}$\,G and neutrino luminosity $L_{\bar{\nu}_{\rm e}}$ in units of $10^{51}$\,ergs s$^{-1}$. All the calculations shown here have the electron fraction $Y_{\rm e}$ of the initial composition in SkyNet nucleosynthesis calculations set to 0.485 \citep{Hoffman1996_2, Pruet2006}. The EOS used in the MHD simulation is the HEOS and the resolution is $256\times 128$ for all the calculations shown in this plot.} 
\label{Mo_isotopes}
\end{figure}

We also explore the possibility for synthesizing isotopes of Molybdenum (Mo), specifically $^{92}\rm Mo$, in neutron-rich winds. Figure \ref{Mo_isotopes} shows the production factor of various isotopes of Molybdenum, where the production factor $P$ is defined as the ratio of abundance of a given isotope in the nucleosynthesis calculation to that of the Solar abundance of the same isotope, 
\begin{equation}
    \label{prod_factor}
    P=Y^{i}/Y^{i}_{\odot}. 
\end{equation}
In Figure \ref{Mo_isotopes}, we normalize $P$ of each isotope to the maximum production factor $P_{\rm max}$ of the isotopes of a given element (Mo in this case). The calculations shown in the plot assume $Y_{\rm e} =$ 0.485 \citep{Hoffman1996_2, Pruet2006}. In the weakly magnetized wind model which exhibits no plasmoid emission ($B_0=5\times 10^{14}$\,G; $L_{\bar{\nu}_{\rm e}}=8\times10^{51}$\,ergs s$^{-1}$), we see that $^{92}\rm Mo$ is the dominant isotope of Mo produced. For more magnetically dominated winds shown in Figure \ref{Mo_isotopes}, $^{92}\rm Mo$ no longer dominates. This demonstrates that $^{92}\rm Mo$ can be synthesized in large quantities in those polar regions of the PNS wind outside the equatorial region from where plasmoids are ejected. We return to the implications of this result for Galactic chemical evolution in Section \ref{discussion}. Similar to \citet{Hoffman1996_2}, we do not find the abundant production of $^{94}\rm Mo$, $^{96,98}\rm Ru$, and $^{102}\rm Pd$ in winds that efficiently synthesize $^{92}\rm Mo$.           

\section{Discussion and Conclusions}
\label{discussion}

We have studied heavy element nucleosynthesis in magnetized rotating proto-neutron star (PNS) winds by post-processing tracer particles embedded into axisymmetric MHD simulations using a nuclear reaction network \citep{Lippuner2017}. We explore PNS wind models across a range of magnetic field strengths, rotation rates, neutrino luminosities, and assumptions about the wind electron fraction (which depends on the details of the neutrino luminosities and spectra).  To the best of our knowledge, this is the first work of its kind.  

Tables \ref{table1} and \ref{table2} summarize the results of our nucleosynthesis calculations across a suite of models.  Although we confirm that unmagnetized PNS winds fail to produce a robust $r-$process extending beyond the third peak (e.g., \citealt{QW1996}), all of our magnetized solutions generate a robust $r-$process extending beyond the third peak (see middle panel of Figure \ref{skynet_B_comp} and Figure \ref{abund_all_plots}).  This successful $r-$process occurs as the result of periodic eruptions of high entropy plasmoids from the closed equatorial zone of the PNS magnetosphere (Figure \ref{plasmoids_sequence}; see also \citealt{Thompson2003,Thompson2018,Desai2023,Prasanna2024}). Combined with rapid expansion, these high entropy conditions provide favorable conditions to achieve a robust $r-$process extending up to the third-peak at $\rm A \sim 195$ via the alpha-rich freeze-out mechanism \citep{Hoffman1997}.

For our fiducial low resolution (LR) simulation of a non-rotating PNS with polar magnetic field $B_0=4\times 10^{15}$\,G using the Helmholtz EOS and considering entropy enhancement from all sources ($\dot{q}_{\rm ext} \ {\rm mode}=H_{\rm all}$ in Tables \ref{table1} and \ref{table2}), we find that the mass flux of elements with mass number $\rm A>190$ is $1.1\times 10^{-6}$\,M$_{\odot}$ s$^{-1}$ (see Table \ref{table2}). The neutrino luminosities (see Section \ref{results_section} for details on neutrino luminosity considered in this paper) considered in this simulation occur $\sim 1-2$\,s after the onset of the PNS cooling phase and lasts for $\sim 1-2$\,s \citep{Pons1999, Roberts2012, Vartanyan2023}. This translates to a production of total mass of $\sim 1-2\times 10^{-6}$\,M$_{\odot}$ of elements in the third $r-$process peak in the first $\sim 1-2$\,s of the cooling phase. We find similar numbers at $B_0=3\times 10^{15}$\,G in the high neutrino luminosity limit considering the total entropy ($\dot{q}_{\rm ext} \ {\rm mode}=H_{\rm all}$; see Table \ref{table1} and Figure \ref{abund_A190}). Due to the threshold effect described in Section \ref{results_section}, we find that even a small reduction in the value of electron fraction used to set the initial composition in the nucleosynthesis calculations leads to significant increase in the production of elements in the third $r-$process peak. At a lower value of neutrino luminosity $L_{\bar{\nu}_{\rm e}}=4\times 10^{51}$\,ergs s$^{-1}$ which occurs $\sim 5-6$\,s after the onset of the cooling phase, we again find a mass flux of $\sim 1-3\times 10^{-6}$\,M$_{\odot}$ s$^{-1}$ of elements with $\rm A>190$ (see Tables \ref{table1} and \ref{table2} and Figure \ref{abund_A190}). These estimates suggests that we can expect $\sim 10^{-5}$\,M$_{\odot}$ of elements with $\rm A>190$ to be produced by a magnetar in the first $\sim 10$\,s of the cooling phase at $B_0\gtrsim 3\times 10^{15}$\,G if we consider entropy enhancement from all sources. 

The PNS wind is non-relativistic during the early cooling phase, but transitions to a relativistic regime as the PNS cools and the neutrino luminosities and energies decrease. Since we use non-relativistic physics in our MHD simulations, we are not able to explore nucleosynthesis as the wind transitions into relativistic regimes. This transition occurs at different times during the cooling phase depending on the magnetic field $B_0$ of the magnetar. For $B_0\sim 3-4\times 10^{15}$\,G, the transition to the relativistic regime occurs $\sim 7-8$\,s after the onset of the cooling phase. For lower values of $B_0\sim 5\times 10^{14}$\,G, the wind transitions to the relativistic regime $\sim 15-20$\,s after the cooling phase begins. In a future work, we plan to incorporate relativistic effects in the MHD simulations and explore nucleosynthesis throughout the cooling phase of the magnetar for various values of $B_0$. 

The Solar mass fractions of $r-$process elements with mass number $\rm A>190$, $\rm A>130$, and $\rm A>100$ are $\sim 10^{-8}$, $\sim 5\times 10^{-8}$, and $\sim 7\times 10^{-8}-10^{-7}$ respectively \citep{Qian2000, Arnould2007, Lodders2020}. For a Galactic mass $M_{\rm G} \sim 10^{11}$\,$\rm M_{\odot}$ in stars and gas \citep{Qian2000}, and a Galactic age $t_{\rm G}\sim 10^{10}$\,yr, the average production rate of $r-$process elements with $\rm A>190$, $\rm A>130$, and $\rm A>100$ is respectively $10^{-7}$\,M$_{\odot}$ yr$^{-1}$, $5\times 10^{-7}$\,M$_{\odot}$ yr$^{-1}$, and $7\times 10^{-7}-10^{-6}$\,M$_{\odot}$ yr$^{-1}$, when averaged over the Galactic history. Given a Galactic supernova rate of $\sim 0.02 \ \rm yr^{-1}$ and a fraction $\sim 40-50\%$ of all neutron stars being magnetars \citep{Beniamini2019}, based on the estimates in the previous paragraph that a single magnetar with $B_0\gtrsim 3\times 10^{15}$\,G can produce $\gtrsim 10^{-5}$\,M$_{\odot}$ of elements with $\rm A>190$ during the cooling phase, we can expect such magnetars to produce $\sim 10^{-7}$\,M$_{\odot}$ yr$^{-1}$ of elements with $\rm A>190$ (assuming all magnetars are born with such surface magnetic fields). This estimate is very close to the production rate of all the heavy $r-$process elements with $\rm A>190$.

We also consider winds from a magnetar at $B_0=10^{15}$\,G. At this value of $B_0$, we do not find high entropy plasmoid ejections that are capable of producing the elements in the third $r-$process peak in the first $\sim 5$\,s of the cooling phase (see \citealt{Prasanna2024}). We present results from a calculation at neutrino luminosity $L_{\bar{\nu}_{\rm e}}=3\times 10^{51}$\,ergs s$^{-1}$ and $L_{\bar{\nu}_{\rm e}}=2\times 10^{51}$\,ergs s$^{-1}$, representative of $\sim 5-7$\,s and $\sim 8-10$\,s after the onset of the cooling phase respectively. From Table \ref{table1} and Figure \ref{abund_A190}, we estimate that magnetars with $B_0=10^{15}$\,G can produce at least $\sim 1-2\times10^{-6}$\,M$_{\odot}$ of elements with $\rm A>190$ in the first $\sim 10$\,s of the cooling phase considering entropy enhancement from all sources ($\dot{q}_{\rm ext} \ {\rm mode}=H_{\rm all}$). As mentioned earlier, as the magnetar cools beyond these luminosities, the wind transitions into a relativistic regime which we cannot currently explore.

To show that production of the heavy $r-$process elements is generic to magnetar birth, we consider winds from magnetars at $B_0=5\times 10^{14}$\,G and neutrino luminosities that occur $\sim 15-20$\,s after the onset of the cooling phase. From Table \ref{table1}, we estimate that such magnetars can produce at least $\sim 2-4\times 10^{-7}$\,M$_{\odot}$ of elements with $\rm A>190$ during the first $\sim 15-20$\,s of the cooling phase considering entropy enhancement from all sources ($\dot{q}_{\rm ext} \ {\rm mode}=H_{\rm all}$).  

From the above estimates, we find that if all Galactic magnetars are born with $B_0\sim 5\times 10^{14}-10^{15}$\,G, they together can account for at least $\sim 5-20\%$ of the Galactic budget of heavy $r-$process elements (including $\rm A>190$ elements in the third peak). On the other hand, if all magnetars were to be born with $B_0\gtrsim 3\times 10^{15}$\,G, they together can account for the entire Galactic budget of heavy $r-$process elements (including $\rm A>190$ elements in the third peak). All of these estimates consider entropy enhancement of the wind material from all sources, denoted by $\dot{q}_{\rm ext} \ {\rm mode}=H_{\rm all}$ in Tables \ref{table1} and \ref{table2}. As we have emphasized above, these estimates consider only the non-relativistic wind phase of magnetars. More heavy $r-$process elements can potentially be produced during the relativistic wind phase, which we intend to explore in a future work. 

Although magnetars can produce a significant fraction of the Galactic heavy $r-$process elements, there is an issue of large overproduction compared to the Solar abundance of elements with $\rm A\lesssim 120$ in magnetar winds due to low entropy wind material off the equatorial region (see Section \ref{results_section} and Figure \ref{abund_all_plots}). This overproduction in our calculations could be a result of incorrect modeling of the electron fraction (a consequence of incorrect modeling of neutrino luminosities and energies) in the regions outside the equatorial region. More speculatively, it is possible that some of the low entropy wind material that overproduces elements with $\rm A\lesssim 120$ is returned to the star through convective downturn, while the high entropy wind material containing the heavy $r-$process elements is ejected \citep{Woosley1994}. In a future work, we intend to study this issue in detail.   

As we emphasize in our simulations (see Section \ref{tracer_paths_sub}), tracers gain entropy from neutrino heating as well as magnetic energy dissipation. Since we do not have a physical model for the current sheet, we present results from nucleosynthesis calculations that consider entropy gain from neutrino heating only (denoted by $\dot{q}_{\rm ext} \ {\rm mode}=H_{\nu}$ in Tables \ref{table1} and \ref{table2}) as well as calculations that take into account the total entropy (denoted by $\dot{q}_{\rm ext} \ {\rm mode}=H_{\rm all}$ in Tables \ref{table1} and \ref{table2}). In the previous paragraphs, we have given estimates for the yield of the heavy $r-$process elements from magnetars considering entropy enhancement of the magnetar wind material from all sources (mode $H_{\rm all}$). In what follows, we estimate yields of $r-$process elements considering enhancement of magnetar wind entropy from neutrino heating only (mode $H_{\nu}$). This is a rather pessimistic estimate because magnetic energy dissipation during magnetic reconnection is sure to increase the entropy of wind material in the plasmoids. But we provide these estimates to get an idea of the lower limit on the yield of $r-$process elements from magnetar winds.   

From Table \ref{table2}, we find that results from the nucleosynthesis calculations in mode $H_{\nu}$ seem to depend significantly on resolution. We find that the entropy of the wind material is sensitive to resolution, with entropy decreasing as the number of radial and angular zones in the simulation increase (see Section \ref{resolution_sub}, Figures \ref{mdot_zeta}, \ref{entr_bins_hydro}, and \ref{rescomp_B4e15_QdotOnly}). In the highest resolution simulations in this paper, we find that entropy from neutrino heating is just around the threshold for a successful $r-$process that produces the third peak, which leads to drastically different results for various resolutions. We discuss the factors that may lead to this difference and convergence in Section \ref{resolution_sub}.   
For the estimates that follow, we use the results from the resolution of our fiducial calculation (256 radial zones and 128 angular zones) assuming that convergence will follow for higher resolutions as well, once we are past the threshold effect of entropy (at larger magnetic fields and/or lower neutrino luminosities or lower values of electron fraction $Y_{\rm e}$).

In our fiducial calculation of a non-rotating PNS at $B_0=4\times 10^{15}$\,G, we find that the mass flux of elements with $\rm A>190$ in mode $H_{\nu}$ is $\dot{M}_{\rm A>190}=3.4\times 10^{-8}$\,M$_{\odot}$ s$^{-1}$ (LR calculation run with the HEOS and marked with ``avg'' in the $Y_{\rm e}$ column in Table \ref{table2}). Decreasing $Y_{\rm e}$ in the initial composition set in the nuclear reaction network SkyNet to 0.42, we find that $\dot{M}_{\rm A>190}=1.2\times 10^{-6}$\,M$_{\odot}$ s$^{-1}$, again pointing to the threshold effect of entropy. For the similar calculations at $B_0=3\times 10^{15}$\,G and neutrino luminosity $L_{\bar{\nu}_{\rm e}}=8\times 10^{51}$\,ergs s$^{-1}$ and $Y_{\rm e}=0.42$, we have $\dot{M}_{\rm A>190}=1.8\times 10^{-7}$\,M$_{\odot}$ s$^{-1}$, while at $L_{\bar{\nu}_{\rm e}}=4\times 10^{51}$\,ergs s$^{-1}$, we have $\dot{M}_{\rm A>190}=3.8\times 10^{-7}$\,M$_{\odot}$ s$^{-1}$ (see Table \ref{table1}). We emphasize that similar values of $\dot{M}_{\rm A>190}$ are possible at higher values of $Y_{\rm e} \sim 0.45$ at lower neutrino luminosities and/or higher magnetic fields. Based on these arguments, we can estimate that magnetars at $B_0\gtrsim 3\times 10^{15}$\,G can release $\sim 4\times 10^{-6}$\,M$_{\odot}$ of material with $\rm A>190$ during the first $\sim 10$\,s of the cooling phase considering entropy from neutrino heating alone. Being conservative and assuming that these results are an overestimate due to the low resolution of $256\times 128$ in the corresponding MHD simulations, we can expect at least a total mass of $10^{-6}$\,M$_{\odot}$ of material with $\rm A>190$ (although this needs to be verified for convergence at higher resolutions than that considered in this paper, and/or with a better modeling of the current sheet). 

Assuming all magnetars in the Galaxy are born with polar magnetic field $B_0\gtrsim 3\times 10^{15}$\,G, they together can contribute at least $\sim 10\%$ of the total Galactic mass of $r-$process elements with $\rm A>190$ (we reiterate that this estimate considers entropy gain from neutrino heating only). We do not find a robust $r-$process extending beyond the third peak at $B_0\le 10^{15}$\,G for the  neutrino luminosities we explore in the non-relativistic limit in this paper when we consider magnetar wind entropy enhancement from neutrino heating only (mode $H_{\nu}$, see Tables \ref{table1} and \ref{table2}). We note that at lower neutrino luminosities than those considered in this paper, as the magnetar wind transitions to the relativistic regime, a robust $r-$process extending beyond the third peak is possible in mode $H_{\nu}$ for $B_0\le 10^{15}$\,G. We reemphasize that the estimates considering entropy from neutrino heating only are definitely underestimates of nucleosynthesis because we are certain that magnetic dissipation during reconnection enhances entropy to a certain extent.    

Although we save the study of nucleosynthesis in proton-rich winds to a future work, we study the prospects for production of a $p-$isotope, $^{92}\rm Mo$, in neutron-rich winds \citep{Hoffman1996_2, Pruet2006}. In Figure \ref{Mo_isotopes} we show the production factor (see equation \ref{prod_factor} and Section \ref{results_section}) of isotopes of Molybdenum. We find from Figure \ref{Mo_isotopes} and our calculations that $^{92}\rm Mo$ can be synthesized in abundant quantities in the regions of the PNS wind outside the high entropy plasmoids. We find that the mass fraction of $^{92}\rm Mo$ in our nucleosynthesis calculations is $\sim 2-3\times 10^{-5}$. Throughout the cooling phase, a PNS ejects $\sim 10^{-3}-10^{-2}$\,M$_{\odot}$ of material, 90-95\% of which does not contain material with enhanced entropy due to plasmoids. We thus can estimate that $\sim 2\times 10^{-8}-2\times 10^{-7}$\,M$_{\odot}$ of $^{92}\rm Mo$ can be produced by PNSs during the cooling phase (assuming the PNS wind remains neutron-rich throguhout the cooling phase). The Solar mass fraction of $^{92}\rm Mo$ is $9.7\times 10^{-10}$ \citep{Lodders2021}. This means that the average production rate of $^{92}\rm Mo$ in the Galaxy is $\sim 10^{-8}$\,M$_{\odot}$ yr$^{-1}$. Given a Galactic supernova rate of $\sim 0.02$\,yr$^{-1}$ and each PNS producing $\sim 2\times 10^{-8}-2\times 10^{-7}$\,M$_{\odot}$ of $^{92}\rm Mo$, we can estimate that PNSs on an average can produce $\sim 4\times 10^{-10}-4\times 10^{-9}$\,M$_{\odot}$ yr$^{-1}$ of $^{92}\rm Mo$. Neutron-rich PNS winds can thus produce $\sim 4-40\%$ of the Galactic abundance of $^{92}\rm Mo$. Similar to the findings in \cite{Hoffman1996_2}, we do not find abundant production of $^{94}\rm Mo$, $^{96,98}\rm Ru$, and $^{102}\rm Pd$ in winds that efficiently synthesize $^{92}\rm Mo$.           

Overall, our simulations and nucleosynthesis calculations suggest that magnetar winds can produce significant quantities of the $r-$process elements in the second and third peaks, but overproduce elements with mass number $\rm A\lesssim120$. Coupled with a recent finding that magnetar giant flares can produce heavy $r-$process elements \citep{Cehula2024, Patel2025, Patel2025_2}, magnetars are a promising source that may have produced a significant fraction of the $r-$process elements in the Universe.  

As a brief summary, the key findings in this work are as follows:
\begin{itemize}
    \item At sufficiently strong magnetar strength magnetic fields, high entropy plasma is quasi-periodically ejected from the closed zone of the magnetar magnetosphere with favorable conditions to produce a robust $r-$process. 
    \item Production of heavy $r-$process elements beyond the third peak ($\rm A>190$) is generic to neutron-rich magnetar winds, occurring even at modest magnetar magnetic fields $\sim 5\times 10^{14}$\,G.
    \item We estimate that magnetars, depending on their surface magnetic fields at birth, can account for $\sim 5-100\%$ of the Galactic heavy $r-$process inventory (including the third peak elements). Such a robust $r-$process in our calculations is accompanied by a large overproduction of elements with $\rm A \lesssim 120$.
    \item We find that neutron rich neutron star winds can account for $\sim 4-40\%$ of the isotope $^{92} \rm Mo$ in the Galaxy.
\end{itemize}

In a future work, we intend to extend our MHD simulations to the relativistic regime to study nucleosynthesis in magnetar winds later in the cooling phase than that considered in this paper. This would enable us to more precisely estimate the contribution of magnetars towards $r-$process element enrichment. We also intend to better model the current sheet to better estimate the nucleosynthesis yields. We also plan to study the effects of fluid mixing at various stages of the $r-$process on the final abundance of elements. Although we show that magnetar winds can produce significant quantities of heavy $r-$process elements, the issue of overproduction relative to Solar abundance of elements with $\rm A\lesssim 120$ needs to be studied in detail, which will be a focus of future work. We also intend to study in greater detail the prospects for nucleosynthesis in proton-rich winds.

\section*{Acknowledgments}
\label{section:acknowledgements}
We thank Chao-Chin Yang for providing us the source code that integrates tracer particles with the MHD simulations (private communication). TP and TAT were supported in part by NASA grant 80NSSC20K0531. TP and BDM acknowledge support from the National Science Foundation (grants AST-2406637, AST-2009255), NASA through the ATP (80NSSC22K0807) and Fermi Guest Investigator Program (grant80NSSC22K1574), and the Simons Foundation (grant 727700).  The Flatiron Institute is supported by the Simons Foundation. We have run our simulations on the Ohio supercomputer \citep{OhioSupercomputerCenter1987}. Parts of the results in this work make use of the colormaps in the CMasher package \citep{CMasher}. We also use the Python packages Numpy \citep{Numpy} and Matplotlib \citep{Matplotlib}.

\bibliography{ref}{}
\bibliographystyle{aasjournal}

\end{document}